\documentclass[english,aps,prc,nofootinbib,superscriptaddress,twocolumn,letter]{revtex4}

\usepackage{graphicx}
\usepackage{amsmath,amssymb,bbold}
\usepackage{subfig}
\usepackage{hyperref}
\usepackage{comment}
\usepackage{amsthm,amsmath,amssymb}\usepackage{mathrsfs}
\usepackage{float}
\usepackage{xcolor}
\usepackage{ulem}

\begin{document}



\title{Asymmetric jet shapes with 2D jet tomography}

\author{Yu-Xin Xiao}
\affiliation{Key Laboratory of Quark \& Lepton Physics (MOE) and Institute of Particle Physics, Central China Normal University, Wuhan 430079, China}

\author{Yayun He}

\affiliation{School of Physics and Optoelectronics,
South China University of Technology, Guangzhou 510640, China}

\author{Long-Gang Pang}
\affiliation{Key Laboratory of Quark \& Lepton Physics (MOE) and Institute of Particle Physics, Central China Normal University, Wuhan 430079, China}

\author{Hanzhong Zhang}
\email{zhanghz@mail.ccnu.edu.cn}
\affiliation{Key Laboratory of Quark \& Lepton Physics (MOE) and Institute of Particle Physics, Central China Normal University, Wuhan 430079, China}
\affiliation{Guangdong Provincial Key Laboratory of Nuclear Science, Institute of Quantum Matter, South China Normal University, Guangzhou 510006, China}
\affiliation{Guangdong-Hong Kong Joint Laboratory of Quantum Matter, Southern Nuclear Science Computing Center, South China Normal University, Guangzhou 510006, China}

\author{Xin-Nian Wang}
\email{xnwang@lbl.gov}
\affiliation{Key Laboratory of Quark \& Lepton Physics (MOE) and Institute of Particle Physics, Central China Normal University, Wuhan 430079, China}
\affiliation{Nuclear Science Division, MS 70R0319, Lawrence Berkeley National Laboratory, Berkeley, CA 94720, USA}



\date{\today}

\begin{abstract}
Two-dimensional (2D) jet tomography is a promising tool to study jet medium modification in high-energy heavy-ion collisions. It combines gradient (transverse) and longitudinal jet tomography for selection of events with localized initial jet production positions. It exploits the transverse asymmetry and energy loss that depend, respectively, on the transverse gradient and jet path length inside the quark-gluon plasma (QGP). In this study, we employ the 2D jet tomography to study medium modification of the jet shape of $\gamma$-triggered jets within the linear Boltzmann transport (LBT) model for jet propagation in heavy-ion collisions. Our results show that jets with small transverse asymmetry ($A_N^{\vec{n}}$) or small $\gamma$-jet asymmetry ($x_{J\gamma}=p_T^{\rm jet}/p_T^\gamma$) exhibit a broader jet shape than those with larger $A_N^{\vec{n}}$ or $x_{J\gamma}$, since the former are produced at the center and go through longer path lengths while the later are off-center and close to the surface of the QGP fireball. In events with finite values of $A_N^{\vec{n}}$, jet shapes are asymmetric with respect to the event plane.  Hard partons at the core of the jet are deflected away from the denser region while soft partons from the medium response at large angles flow toward the denser part of QGP.  Future experimental measurements of these asymmetric features of the jet shape can be used to study the transport properties of jets and medium responses. 
\end{abstract}


\maketitle


\section{INTRODUCTION}

A few microseconds after the Big Bang, the Universe was filled with a novel state of matter called quark-gluon plasma (QGP) that consists of deconfined quarks and gluons.  In the past two decades, relativistic heavy-ion collision experiments have been carried out at the Relativistic Heavy Ion Collider (RHIC) at BNL and the Large Hadron Collider (LHC) at CERN to produce such hot and dense QGP matter and study its properties.

Together with the formation of QGP, large transverse momentum partons ($p_T$) are also produced in the early stage of relativistic heavy-ion collisions that will traverse the QGP and encounter multiple scattering with quarks and gluons inside the medium. These scatterings will also induce gluon bremsstrahlung and cause a significant amount of energy loss to the propagating parton, leading to a phenomenon called jet quenching \cite{Gyulassy:1990ye, Wang:1992qdg, Qin:2015srf}. Large transverse momentum ($p_T$) hadrons and jets are significantly suppressed due to jet quenching in heavy-ion collisions, as compared to large $p_T$ hadrons and jets in proton-proton collisions. Jet quenching has been observed in experiments by PHENIX \cite{PHENIX:2010nlr} and STAR \cite{STAR:2002ggv} Collaboration at RHIC, ALICE \cite{ALICE:2012vgf}, ATLAS \cite{ATLAS:2011ah} and CMS \cite{CMS:2012tqw,CMS:2017xgk} Collaboration at LHC. In addition, large $p_T$ dihadrons, $\gamma$-hadrons and dijets are also suppressed and exhibit azimuthal decorrelation \cite{Wang:2002ri, Vitev:2004bh, Zhang:2007ja, Zhang:2009rn, PhysRevLett.104.132001}. Full jet structure is also modified \cite{Qin:2010mn, He:2011pd, Young:2011qx, Barata:2023zqg} by the QGP medium as observed at both RHIC and LHC \cite{STAR:2006vcp, Casalderrey-Solana:2010bet, ALICE:2010khr, ATLAS:2010isq, CMS:2018jco}. 


Theoretical studies have revealed that the strength of jet quenching is proportional to jet transport coefficient $\hat q$, which is defined as the transverse momentum squared per unit length along the parton trajectory \cite{Baier:1996sk, Baier:1996kr, Baier:1998kq, Guo:2000nz, Wang:2001ifa, Casalderrey-Solana:2007xns,Majumder:2009ge}. This coefficient is directly related to the medium gluon density. In high-energy heavy-ion collisions, the spatial distribution of gluon density and the jet transport coefficient are not uniform inside the QGP fireball. They reach their highest values in the central region of the QGP where the temperature is also the highest. The initial spatial distributions are given by the nuclear geometry of two colliding nuclei and the time evolution can be determined from hydrodynamic model of heavy-ion collisions. Jet tomographic techniques have been developed that exploit these non-uniform distributions of the medium and jet quenching strength to reveal information about the locations of the initial jet production.

The longitudinal tomography was first proposed in Refs.~\cite{Zhang:2007ja,Zhang:2009rn} to localize the initial jet production positions along the direction of jet propagation in the study of dihadron and $\gamma$-hadron spectra in heavy-ion collisions. In the case of $\gamma$-jet processes, low $p_T$ $\gamma$-associated hadrons are produced primarily through the fragmentation of $\gamma$-jets created in the central region of the medium (volume emission) after losing some amount of energy. The high $p_T$ $\gamma$-associated hadrons, however,  mainly come from $\gamma$-jets initially produced in the outer corona region (surface emission) that suffer little or no energy loss. 
Furthermore, a significant fraction of the surface emission is tangential to the medium surface. In a more recent study \cite{He:2020iow}, a transverse or gradient jet tomography was developed. In this case, an asymmetry observable $A_N^{\vec{n}}$ caused by the transverse gradient of the medium for high $p_T$ hadrons inside a jet is introduced to localize the initial position of jet production perpendicular to the jet propagation direction. Two-dimensional (2D) jet tomography combines longitudinal and transverse tomography to localize the initial jet production positions in the transverse plane. Such 2D jet tomography has been applied to select special classes of events with localized region of initial jet production so that signals of jet-induced medium response are enhanced \cite{Yang:2021qtl,Yang:2022yfr}. 

In this work, we study medium-modified $\gamma$-jet shape in events of jet production selected with transverse and longitudinal jet tomography in heavy-ion collisions, using the linear Boltzmann transport (LBT) model \cite{Li:2010ts, He:2015pra, Cao:2016gvr, Wang:2013cia, Cao:2020wlm,Luo:2023nsi} for jet transport simulations in the QGP medium. We consider in particular the asymmetric jet shape caused by flow and gradient of the hot QGP medium along the path of jet propagation in events selected with the 2D jet tomography.

The remainder of this paper is organized as follows.
Sec.~\uppercase\expandafter{\romannumeral2} introduces the LBT model and the $\gamma$-jet shape in p+p and Pb+Pb collisions.
Sec.~\uppercase\expandafter{\romannumeral3} describes the technique of transverse and longitudinal tomography for $\gamma$-jet events. The numerical results for the medium-modified jet shape via transverse and longitudinal tomography are presented in Sec.~\uppercase\expandafter{\romannumeral4}. 
In Sec.~\uppercase\expandafter{\romannumeral5}, we investigate the asymmetric jet shape in events with special class of jet propagation path relative to the geometry of the dense QGP medium as selected by 2D jet tomography. A summary is given in Sec.~\uppercase\expandafter{\romannumeral6}.

\section{ Jet shape in the LBT model}

The linear Boltzmann transport (LBT) model \cite{Li:2010ts, He:2015pra, Cao:2016gvr, Wang:2013cia, Cao:2020wlm}  was developed to study jet propagation in quark-gluon plasma in heavy-ion collisions, and to describe not only parton energy loss but also jet-induced medium excitation. It has been used to describe experimental data on suppression of single hadrons \cite{Cao:2016gvr}, single jets \cite{He:2018xjv}, and $\gamma$-hadron \cite{Chen:2017zte} and $\gamma$-jet correlations \cite{Luo:2018pto} in heavy-ion collisions at both RHIC and LHC energies.

The transport of jet shower and recoil partons is described by the linear Boltzmann equations,
\begin{align}
\begin{split}
p_a & \cdot \partial f_a = \int \prod_{i=b,c,d} \frac{d^3 p_i}{2E_i (2\pi)^3} \frac{\gamma_b}{2}(f_c f_d - f_a f_b) |\mathcal{M}_{ab\rightarrow cd}|^2 \\
&\times S_2(\hat{s}, \hat{t}, \hat{u}){2\pi}^4 \delta^4 (p_a + p_b - p_c - p_d) + {\rm inelastic},
\end{split}
\end{align}
where $|\mathcal{M}_{ab\rightarrow cd}|$ is the leading-order elastic scattering amplitude that depends on the Mandelstam variables $\hat{s}, \hat{t}$, and $\hat{u}$. $f_i=(2\pi)^3 \delta^3 (\vec{p}-\vec{p}_i)\delta^3 (\vec{x}-\vec{x_i}-\vec{v_i}t) (i=a,c)$ is the phase space density for jet shower partons before and after scattering.
$f_i=1/(e^{p_i\cdot u/T}\pm 1)(i=b,d)$ is the phase space distribution of thermal partons in the QGP medium with local temperature $T$ and fluid 4-velocity $u$. The regularization factor,
\begin{equation}
S_2(\hat{s}, \hat{t}, \hat{u}) = \theta (\hat{s}\geq 2\mu_D^2)\theta(-\hat{s}+\mu_D^2\leq\hat{t}\leq -\mu_D^2),
\end{equation}
is used to regulate the collinear divergence in $|\mathcal{M}_{ab\rightarrow cd}|$, and $\mu_D^2=3g^2T^2/2$ is the Debye screening mass.

The description of inelastic processes of induced gluon radiation in the LBT model is based on the high-twist approach \cite{Guo:2000nz, Zhang:2003wk} with the radiative gluon spectrum,
\begin{equation}
\frac{dN^a_g}{dzdk^2_{\perp}d\tau} = \frac{6\alpha_s P_a(z)k^4_\perp}{\pi (k^2_\perp + z^2 m^2 )^4} \frac{p\cdot u}{p_0} \hat{q}_a (x) \sin^2 \frac{\tau - \tau_i}{2\tau_f},
\end{equation}
where $m$ is the mass of the propagating parton $a$, $z$ and $k_\perp$ are the energy fraction and transverse momentum of radiated gluon, respectively.
$\alpha_s$ is the strong coupling constant, and $\hat{q}_a$ is the jet transport coefficient, defined as the transverse momentum transfer squared per unit path for the parton $a$ traveling in the medium.
$P_a(z)$ is the splitting function for $a\rightarrow a+g$, and $\tau_f = 2Ez(1-z)/(k^2_\perp + z^2 M^2)$ is the formation time of the radiated gluon \cite{Luo:2018pto}.  In LBT model, recoiled medium partons are also transported according to the Boltzmann equation, the same as jet shower partons, going through further elastic and inelastic scattering once they are produced. The depletion of the medium due to back-reaction is also taken into account through ``negative" partons whose energy and momentum must be subtracted from the final observables. Recoil and ``negative" partons are generally referred to as medium-response.

In this study, we employ the multi-phase transport (AMPT) model \cite{Lin:2004en} to generate the initial energy density distribution and collision geometry. These include the distribution of binary collisions $N_{\rm coll}$, which is used to determine the initial positions of $\gamma$-jet production, as well as the initial energy density distribution that serves as the initial condition for CLVisc 3+1D viscous hydrodynamic evolution \cite{Pang:2012he, Pang:2018zzo} of the bulk QGP medium. PYTHIA 8 \cite{Sjostrand:2006za, Sjostrand:2007gs} is utilized to generate the momenta of the initial jet shower partons in $\gamma$-jet events.  The LBT model is then employed to describe the propagation and transport of the jet shower partons inside the QGP medium  whose evolution is provided by the CLVisc hydrodynamics with the initial condition given by the AMPT model. The information on local temperature and fluid four-velocity from the CLVisc hydrodynamics is used in LBT to determine the elastic and inelastic scattering rates. The initial time and freeze-out temperature in CLVisc hydrodynamics are set to $\tau_0=0.2$ fm/c and $T_f=137$ MeV, respectively, which also determine the beginning and end of the parton-medium interaction. We use the FastJet package \citep{Cacciari:2011ma, Cacciari:2005hq} with the anti-$k_t$ jet algorithm to reconstruct full jets using the partonic information of the final jets. 

The jet shape is defined as the probability density of transverse momentum distribution inside a jet cone \cite{CMS:2018jco, Luo:2018pto}, 
\begin{equation}
\rho(r) = \frac{1}{\Delta r} \frac{\sum_i^{N_{\rm jet}} p_T^i (r-\frac{1}{2}\Delta r, r+\frac{1}{2}\Delta r)}{\sum_i^{N_{\rm jet}} p_T^i(0, R)},
\label{eqn:shape}
\end{equation}
calculated from the parton distribution inside the reconstructed jets, where the distance $r$ between the track partons and the jet axis is defined as $r = \sqrt{\left( \eta - \eta_{\rm jet} \right)^2 + \left( \phi - \phi_{\rm jet} \right)^2}$ in the plane of pseudo-rapidity $\eta$ and azimuthal angle $\phi$. The total energy from the $i$-th jet in the circular annulus with inner radius $r_1=r - \Delta r / 2$ and outer radius $r_2=r + \Delta r / 2$ is defined as $p_T^i(r_1, r_2)=\sum_{{\rm assoc}\in \Delta r}p_T^{\rm assoc}$ where $\Delta r=r_2 - r_1$ is the width of the annulus. Additionally, the total energy inside the jet cone with a radius $R$ for the $i$-th jet is given by $p_T^i(0, R)$. The summation is over the total number of jet $N_{\rm jet}$ analyzed.

\begin{figure}[htb]
\raggedright
\includegraphics[width=0.48\textwidth]{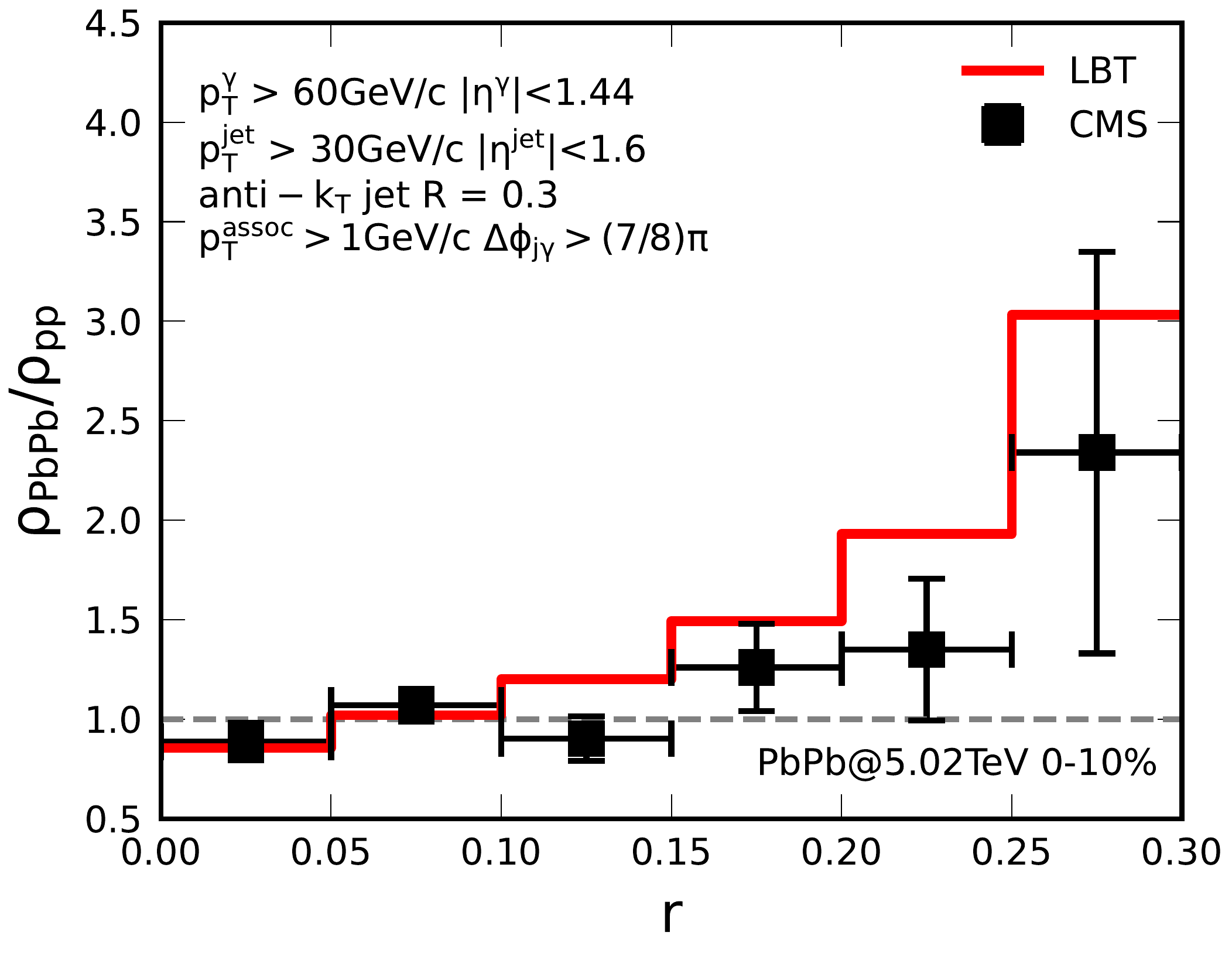}
\captionsetup{justification=raggedright}
\caption{(Color online) The ratio of the $\gamma$-jet shape between 0-10\% Pb+Pb and p+p collisions as a function of $r$ at $\sqrt{s_{NN}} = 5.02$ TeV, from LBT calculations (red-solid) as compared with the CMS data (black-squares) \cite{CMS:2018jco}.}
\label{obser}
\end{figure}

In the following calculations, we select the trigger photon with $p_T^\gamma>$ 60 GeV/$c$ in Pb+Pb collisions at $\sqrt{s_{NN}} = 5.02$ TeV. The cone size of the $\gamma$-triggered jets is set as $R=0.3$, and the lower threshold of transverse momentum for the jets is set at $p_T^{\rm jet} > 30$ GeV/c, with associated partons selected with $p_T^{\rm assoc}>1$ GeV/$c$. The pseudorapidities for the $\gamma$ and jets are constrained within $|\eta_{\gamma}|<1.44$ and $|\eta_{\rm jet}|<1.6$, and their azimuthal angle differences $|\Delta\phi_{j\gamma}|$ are restricted to be larger than $(7/8)\pi $. 

Shown in Fig. \ref{obser} is the ratio of $\gamma$-jet shape in 0-10\% Pb+Pb over that in p+p collisions as a function of $r$ from the LBT model simulations as compared with the CMS data \cite{CMS:2018jco}. The LBT result is in good agreement with experimental data. The numerical ratio indicates an enhancement at large r region, suggesting that a significant fraction of the transverse momentum is transported to larger angles with respect to the jet axis in Pb+Pb relative to p+p collisions.  This is partially due to radiated gluons, but most of the enhancement is caused by jet-induced medium response at large angles \cite{Luo:2018pto}.  This medium modification of the jet shape can also be described by the coupled LBT hydro model \cite{Chen:2017zte,Yang:2021qtl,Yang:2022nei} where the soft mode of jet-induced medium-response is modeled through hydrodynamics that couples to the Boltzmann transport of hard partons.

\section{Transverse and longitudinal jet tomography of heavy-ion collisions}

The path-length dependence of the parton energy loss during the jet propagation within the QGP medium can be utilized to localize the initial jet production positions along the longitudinal direction, a technique known as longitudinal jet tomography \cite{Zhang:2009rn}. Additionally, non-zero spatial gradients of the parton transport coefficient $\hat{q}$ perpendicular to their propagation directions arise due to temperature gradients in the QGP fireball. Parton propagation in such a non-uniform medium  leads to an asymmetrical transverse momentum distribution \cite{He:2020iow,Fu:2022idl,Barata:2022utc,Barata:2023qds} of final state hadrons. In a class of events with a given transverse asymmetry, a jet observable defined to characterize the asymmetric transverse momentum distribution of hadrons or partons inside a jet,  the distribution of the initial jet transverse positions is found localized in a region with a corresponding transverse gradient of the jet transport coefficient. The average transverse position of the initial jet production has a monotonic dependence on the transverse asymmetry, therefore, enabling transverse jet tomography. By considering both longitudinal and transverse jet tomography, one is able to localize the jet production positions in the transverse plane, resulting in a 2D jet tomography \cite{Yang:2021qtl,Yang:2022yfr}.

\begin{figure}[htpb]
\centering
\includegraphics[width=0.35\textwidth]{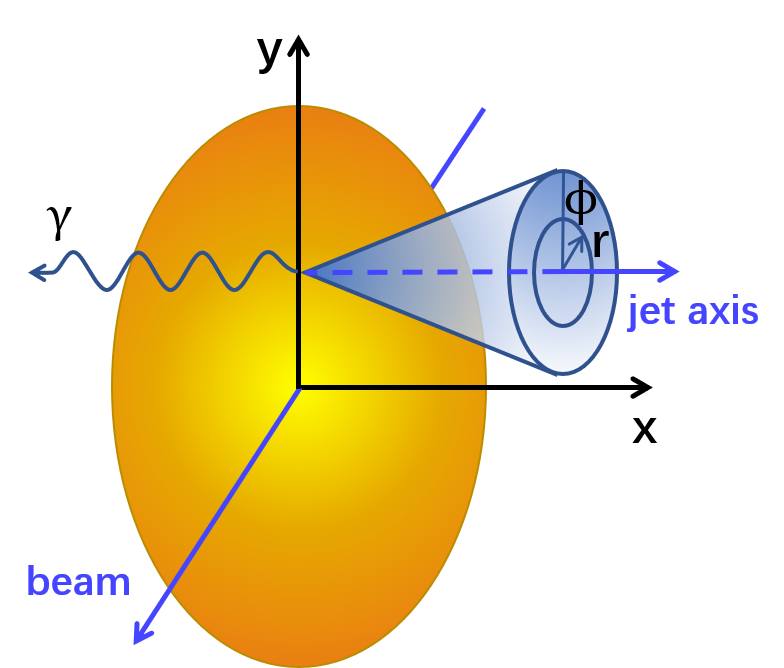}
\captionsetup{justification=raggedright}
\caption{(Color online) Illustration of the $\gamma$-jet configuration relative to the event plane and the azimuthal angle distribution of the jet shape inside the jet cone.}
\label{illustration}
\end{figure}

 In our study here, we  consider the event plane along the $x$-axis and select the direction of the trigger photon to be along the negative $x$-axis, while the correlated jets propagate approximately along the positive $x$-axis as illustrated in Fig.~\ref{illustration}.  The asymmetry $A_N^{\vec{n}}$ used for transverse jet tomography is defined as \cite{He:2020iow},
\begin{equation}
A_N^{\vec{n}} = \frac{\sum_a\int d^3 r d^3 k f_a\left( \vec{k}, \vec{r} \right) {\rm Sign} \left( \vec{k} \cdot \vec{n} \right)}{\sum_a\int d^3 r d^3 k f_a\left( \vec{k}, \vec{r} \right)},
\label{eqn:asym}
\end{equation}
where $f_a$ represents the phase-space distribution of partons (hadrons),  $\vec{n}$ denotes the normal direction of plane defined by the beam direction and the direction of the trigger particle, which is the same as the positive $y$-axis in our set-up in  Fig.~\ref{illustration}. The summation is over all partons (hadrons) inside a jet.

\begin{figure*}[htb]
\centering
\includegraphics[width=0.75\textwidth]{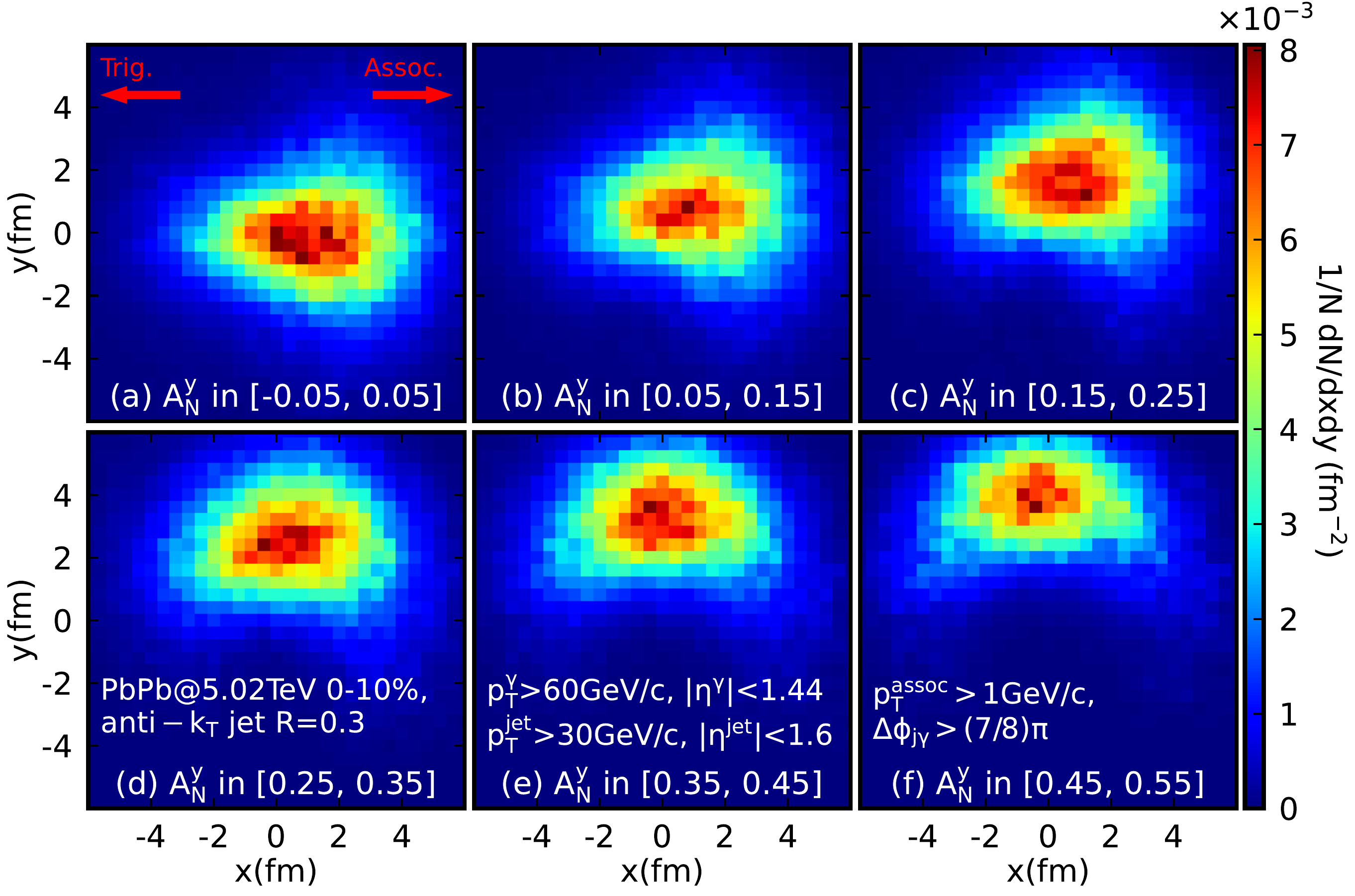}
\captionsetup{justification=raggedright}
\caption{(Color online). The $\gamma$-jet production rates $1/N d^2N/dxdy$ as a function of initial jet production positions ($x$, $y$) in the transverse plane in 0-10\% Pb+Pb collisions at 5.02 TeV, selected with 6 different ranges of transverse asymmetry, $A_N^y$ = [-0.05, 0.05], [0.05, 0.15], [0.15, 0.25], [0.25, 0.35], [0.35, 0.45] and [0.45, 0.55] in the 6 panels, respectively. The trigger $\gamma$ is selected along the negative $x$-axis direction with $p_T^{\gamma}>$ 60 GeV/c and $p_T^{\rm jet}>$ 30 GeV/c. Other kinetic cuts are the same as in Fig. \ref{obser}.}
\label{asymmetry}
\end{figure*}

Shown in Fig.~\ref{asymmetry} are the final $\gamma$-jet production rates $(1/N) d^2N/dxdy$ as a function of the initial locations ($x$, $y$) in the transverse plane in 0-10\% Pb+Pb collisions at 5.02 TeV with different values of the jet transverse asymmetry $A_N^{\vec{n}}$. The six panels, (a), (b), (c), (d), (e), and (f),  correspond to 6 different bins of the transverse asymmetry, $A_N^y$ = [-0.05, 0.05], [0.05, 0.15], [0.15, 0.25], [0.25, 0.35], [0.35, 0.45] and [0.45, 0.55], respectively. As we can see from the numerical results, the typical initial jet production transverse position ($y$) shifts from the center to the outer region of the medium when $A_N^{\vec{n}}$ (or $A_N^{y}$) is increased from [-0.05, 0.05] to [0.45, 0.55].
For small values of $A_N^{y}$, $\gamma$-jets originate from the center region and jets have to traverse the whole volume of the QGP as depicted in panel (a). Conversely, for large values of $A_N^{y}$, $\gamma$-jets emerge from surface areas of the QGP and jets propagate tangentially to the medium surface, as shown in panel (f). 

It is worth noting that the selected jet production positions also exhibit an asymmetric distribution along the x-axis. This is attributed to the path length dependence of $A_N^{y}$, the parton energy loss and the azimuthal angle restriction. The path-length dependence of the parton energy loss biases the initial production position of the selected jets closer to the surface (positive $x$ values). The probability to have a smaller value of $A_N^{y}$ with a longer path length is also rare as compared to a shorter path length. Additionally, longer path lengths may result in a larger drift of the jet direction, outside the azimuthal cut-off $|\Delta \phi_{j\gamma}| < (7/8)\pi $.

\begin{figure}[htpb]
\raggedright
\includegraphics[width=0.48\textwidth]{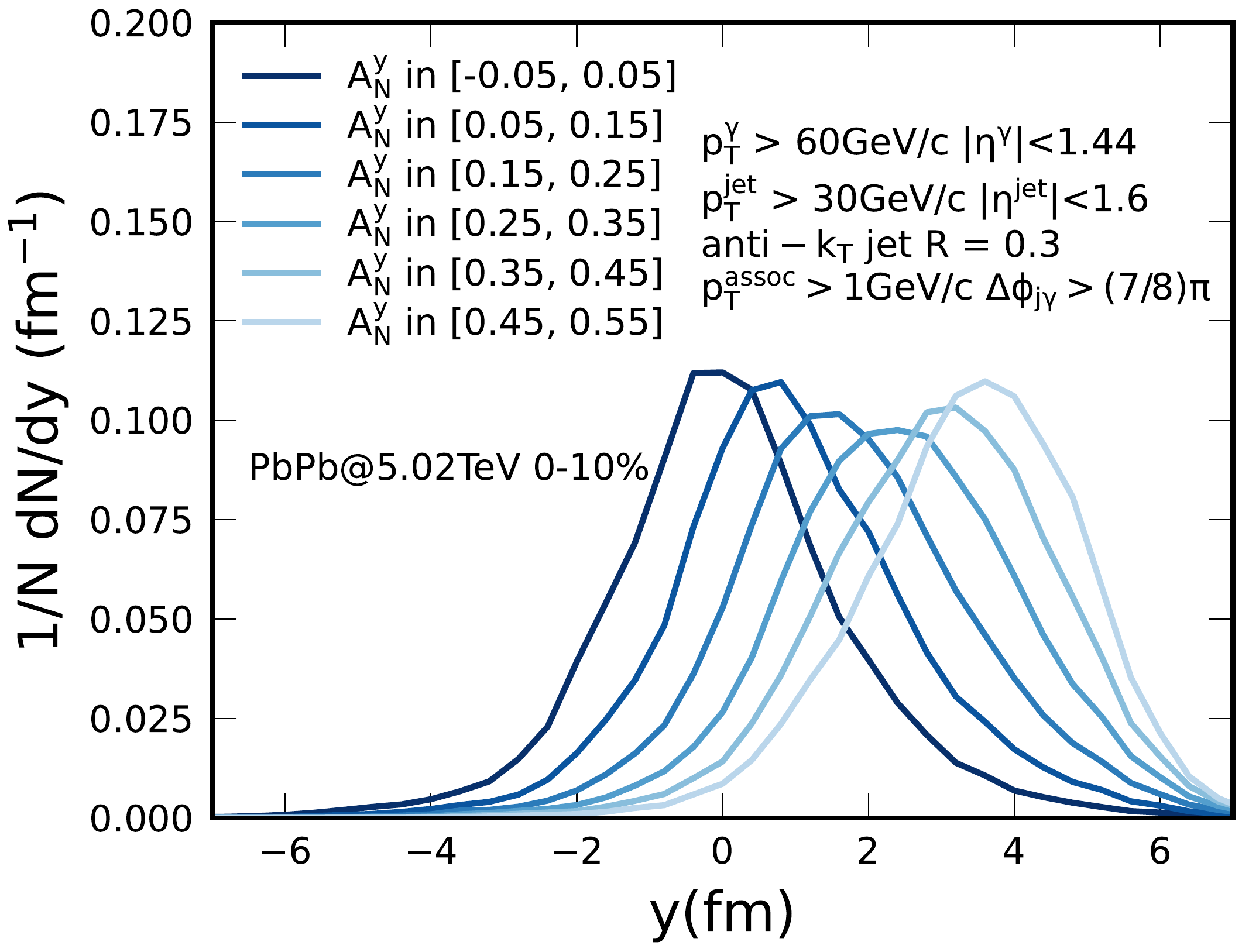}
\captionsetup{justification=raggedright}
\caption{(Color online) The $\gamma$-jet production rates $1/N dN/dy$ as a function of the initial jet production position $y$, selected with different asymmetry $A_N^{y}$. The kinematic cuts are the same as in Fig. \ref{asymmetry}.}
\label{y_distribution}
\end{figure}

The integration of the initial $\gamma$-jet production rates along the $x$-axis for different selected transverse asymmetry $A_N^{y}$ projects the 2-dimensional $\gamma$-jet rate onto the $y$-axis, as shown in Fig.~\ref{y_distribution}. It is clear from this figure that the peak of the transverse distribution of the initial jet production position shifts towards large values of  $y$ with increased value of the transverse asymmetry $A_N^{y}$. The variance appears to be similar across different $A_N^{y}$ ranges. The shape of the approximately asymmetric distribution $(1/N) dN/dy$ does not change significantly when $A_N^{y}$ varies between 0 and 0.35, corresponding to the central regions of QGP. However, the transverse distribution becomes more asymmetric for large $A_N^{y}$ because of the surface bias by the path-length dependence of $A_N^{y}$.  

\begin{figure}[htpb]
\raggedright
\includegraphics[width=0.48\textwidth]{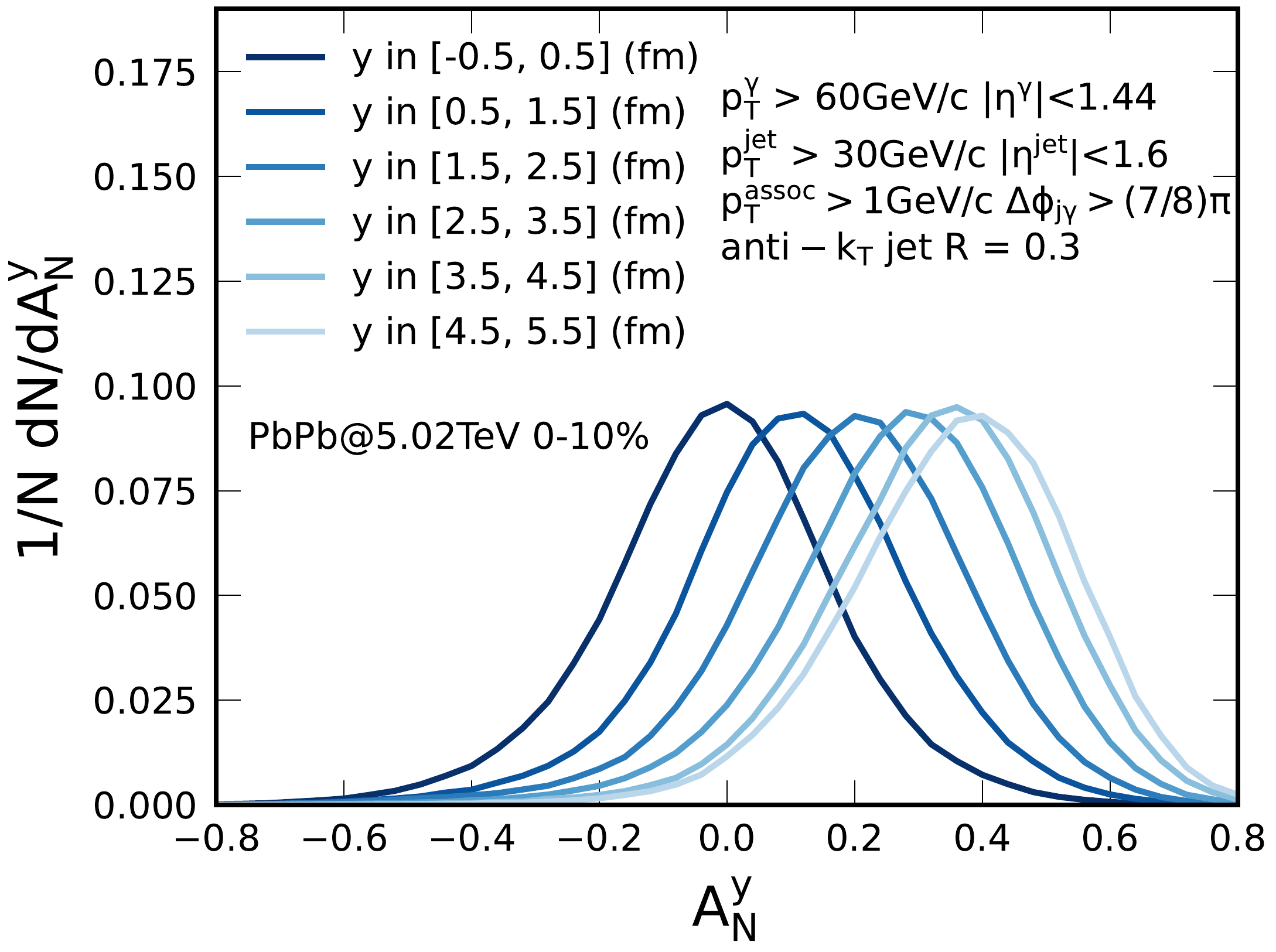}
\captionsetup{justification=raggedright}
\caption{(Color online) The $\gamma$-jet production rates $1/N dN/dA_N^{y}$ as a function of $A_N^{y}$, with given initial jet production position $y$. The corresponding cuts are the same as in Fig. \ref{asymmetry}.}
\label{A_distribution}
\end{figure}

For the purpose of an illustration, one can also select the initial transverse jet production positions of $\gamma$-jets and calculate the corresponding distribution of the transverse asymmetry. Fig.~\ref{A_distribution} shows the $\gamma$-jet production rates $(1/N)dN/dA_N^{y}$ as a function of $A_N^{y}$ with given bins of $y$ coordinate of the initial jet production. With given initial bins of $y$ ranging from $y$ = [-0.5, 0.5] fm to [4.5, 5.5] fm, the peak of the distribution shifts from small to large $A_N^{y}$ values. In summary, according to Figs. \ref{y_distribution} and \ref{A_distribution}, the transverse asymmetry $A_N^{y}$ of the final state particles inside a jet does localize the initial transverse positions of $\gamma$-jet production.

\begin{figure*}[htpb]
\centering
\includegraphics[width=0.7\textwidth]{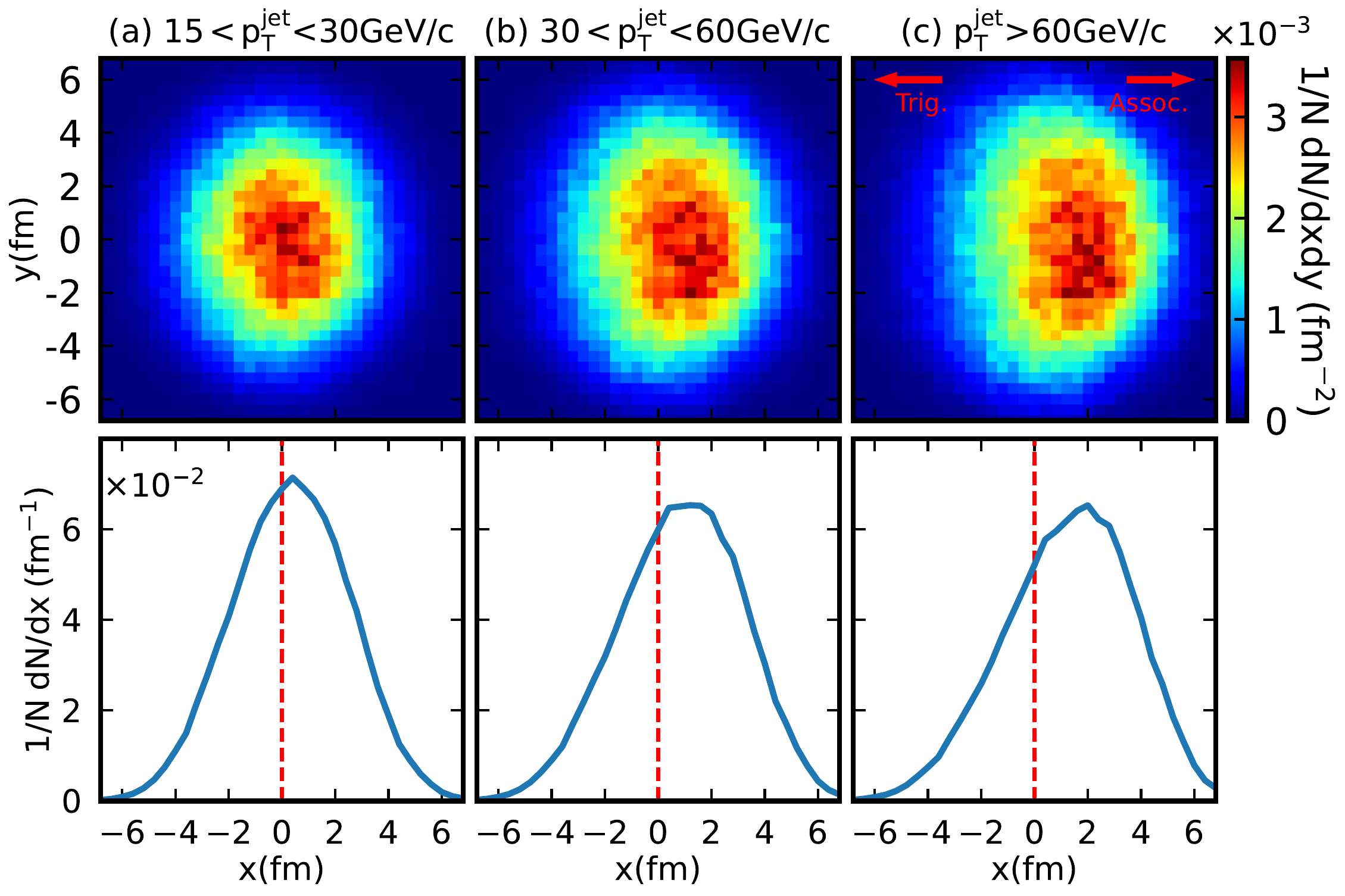}
\captionsetup{justification=raggedright}
\caption{(Color online). (Upper) The $\gamma$-jet production rates $1/N d^2N/dxdy$ as a function of initial jet production locations ($x$, $y$) in the transverse plane in 0-10\% Pb+Pb collisions at 5.02 TeV, selected with 3 different jet transverse momentum, $p_T^{\rm jet}$ = [15, 30] GeV/c, [30, 60] GeV/c and $p_T^{\rm jet}>$ 60 GeV/c, respectively. These jets are triggered by photons with transverse momentum $p_T^{\gamma}>$ 60 GeV/c. Other cuts are the same as in Fig. \ref{asymmetry}. (Lower) The corresponding projections onto the $x$-axis.}
\label{pt_tomography}
\end{figure*}

 To illustrate the power of longitudinal jet tomography in localizing the initial jet production positions, we show in the upper panels of Fig.~\ref{pt_tomography} the contour distributions of the initial $\gamma$-jet production rates $(1/N) d^2N/dxdy$ in the transverse plane ($x$, $y$) for different jet transverse momentum (a) $p_T^{\rm jet}$ = [15, 30] GeV/$c$, (b) $p_T^{\rm jet}$ = [30, 60] GeV/$c$ and (c)  $p_T^{\rm jet}>$ 60 GeV/$c$, respectively, for direct photons with the transverse momentum $p_T^\gamma> 60$ GeV/$c$.
The lower panels give the corresponding projections onto the $x$-axis, $(1/N) dN/dx = (1/N) \int dy (d^2N/dxdy)$.
For fixed photon transverse momentum $p_T^\gamma$, a higher value of $p_T^{\rm jet}$ indicates less jet energy loss due to a shorter path-length of jet propagation. Conversely, smaller $p_T^{\rm jet}$ corresponds to larger jet energy loss and longer path-length. This is indeed seen in Fig.~\ref{pt_tomography}: $\gamma$-jets with smaller $p_T^{\rm jet}$ are primarily produced through volume emissions that undergo long path lengths and have significant energy loss, while $\gamma$-jets with larger $p_T^{\rm jet}$ are mainly produced through surface emissions with shorter path lengths and less energy loss. Varying $p_T^{\rm jet}$ while keeping $p_T^\gamma$ fixed thus provides a method to approximately localize the initial $\gamma$-jet production locations in the longitudinal direction along the jet path.

\section{Jet shape modification with jet tomography}

In this section, we investigate the modification of jet shape for $\gamma$-jets produced in different regions of the QGP fireball as selected by jet tomography. During the jet propagation, hard partons tend to penetrate the dilute region of the medium with low gluon density, while soft recoil partons are more likely to be created in the dense region with high gluon density. By localizing $\gamma$-jets to different regions, we can study the modification of jet shape for jets with different path lengths and temperature gradients. In particular, temperature gradients along the direction perpendicular to the jet propagation may lead to an asymmetric jet shape. We will employ transverse tomography, longitudinal tomography, and a combination of these two methods, namely 2D-tomography, to localize $\gamma$-jets in the transverse plane. The kinetic cuts used in this section remain the same as before.



\subsection{Jet shapes selected with transverse tomography}

\begin{figure}
\raggedright
\includegraphics[width=0.48\textwidth]{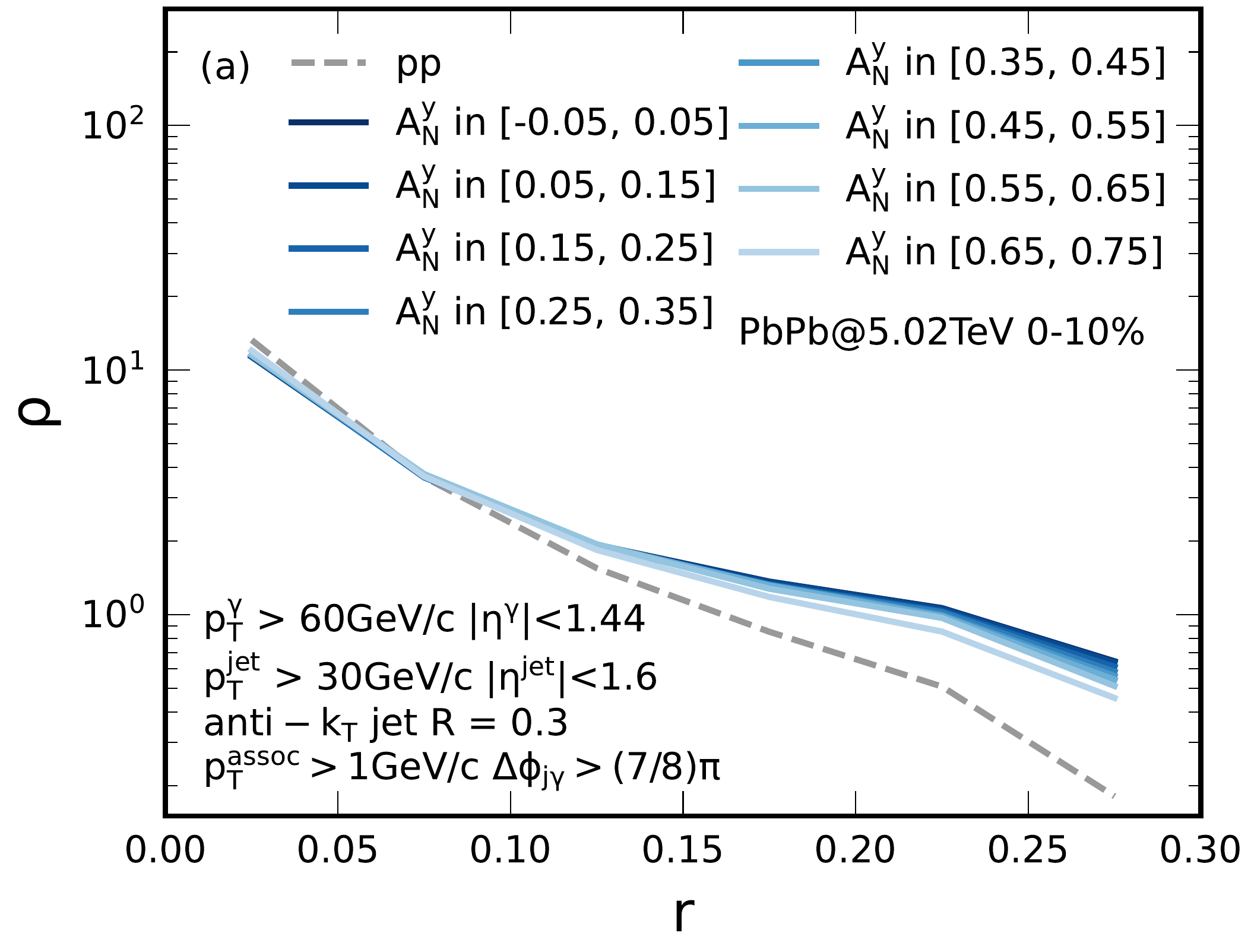}
\includegraphics[width=0.48\textwidth]{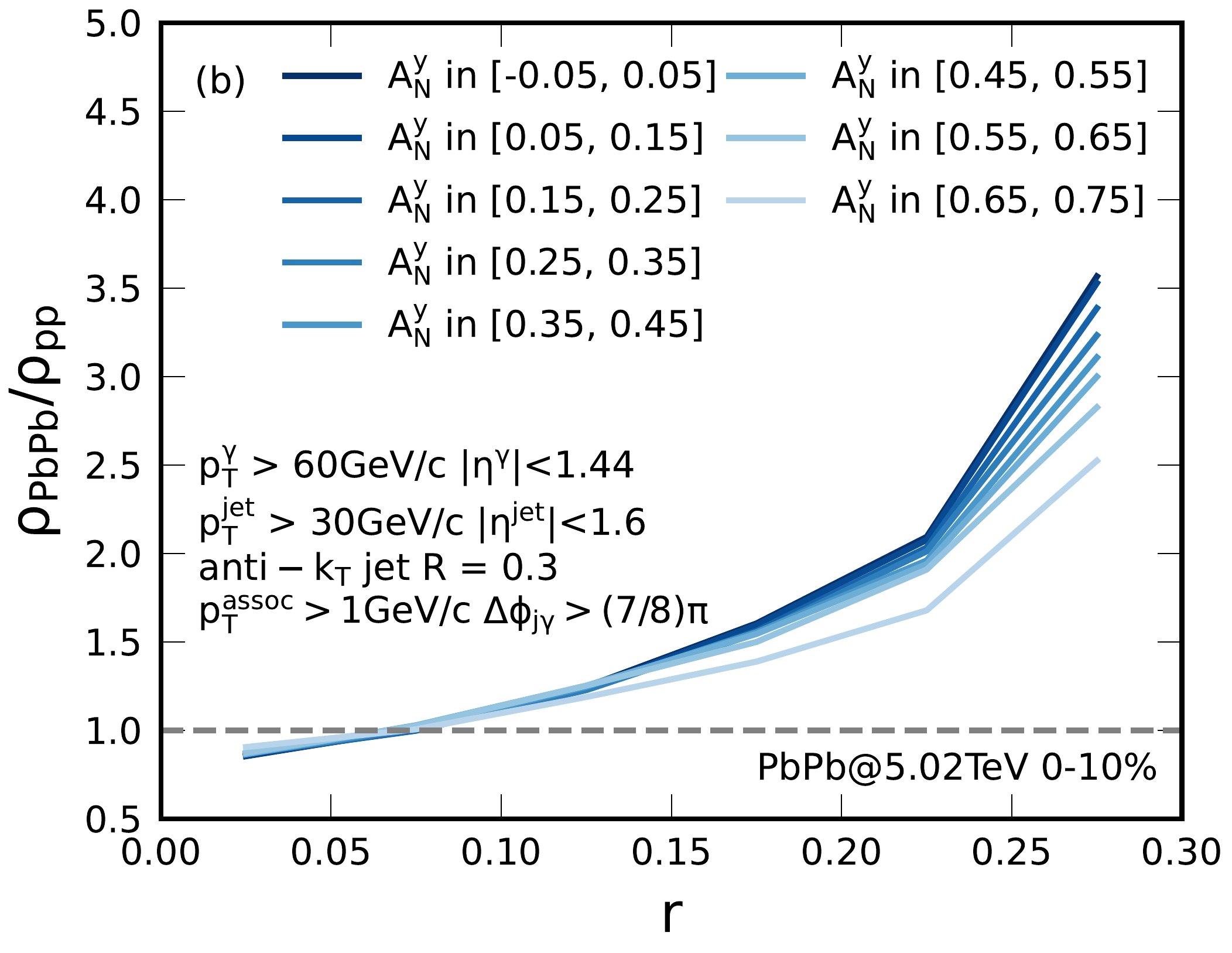}
\captionsetup{justification=raggedright}
\caption{(Color online) 
(a) Jet shape of $\gamma$-jets as a function of $r$ in 0-10\% Pb+Pb (solid curves) selected with different ranges of $A_N^y$ and p+p (dashing curve) collisions at $\sqrt{s_{NN}}=5.02$ TeV. (b) Ratios of jet shapes in Pb+Pb over p+p collisions as a function of $r$.}
\label{fix_asymmetry}
\end{figure}

 We first consider jet events selected with the transverse jet tomography and examine the corresponding modification of the jet shape.  Shown in Fig.~\ref{fix_asymmetry} are (a) the jet shape as a function of the radius $r$ in 0-10\% Pb+Pb collisions (solid curves), for $\gamma$-jets selected with different ranges of the transverse asymmetry $A_N^y$ from [-0.05, 0.05] to [0.65, 0.75], as compared to the jet shape in p+p collisions (dashed line) at $\sqrt{s_{NN}}=5.02$ TeV, and (b) the ratios of jet shapes in Pb+Pb and p+p collisions. In both Pb+Pb and p+p collisions, the jet shape decreases monotonically with the radius $r$, indicating that most of the jet energy is concentrated in the core region close to the jet axis. However, the jet shape in Pb+Pb is in general broader than that in p+p collisions. This suggests that some fraction of the transverse momentum  lost by hard partons at the core of the jet is transported to large angles by soft partons (radiated gluons and medium response) with respect to the jet axis in Pb+Pb relative to p+p collisions. Furthermore, $\gamma$-jets selected with smaller transverse asymmetry $A_N^y$ exhibit a broader jet shape, indicating that the path length is longer and quenching effect is stronger for jets produced in the central region of QGP as localized by smaller values of $A_N^y$. According to the picture we have illustrated in the last section, increasing $A_N^y$ will shift the $\gamma$-jet production positions toward the outer region of QGP. This will reduce the jet energy loss and jet quenching effect with smaller broadening of the jet shape. Therefore, the width of the jet shape in Pb+Pb collisions decreases monotonically as the values of $A_N^y$ increase.

\begin{figure}[ht]
\raggedright
\includegraphics[width=0.48\textwidth]{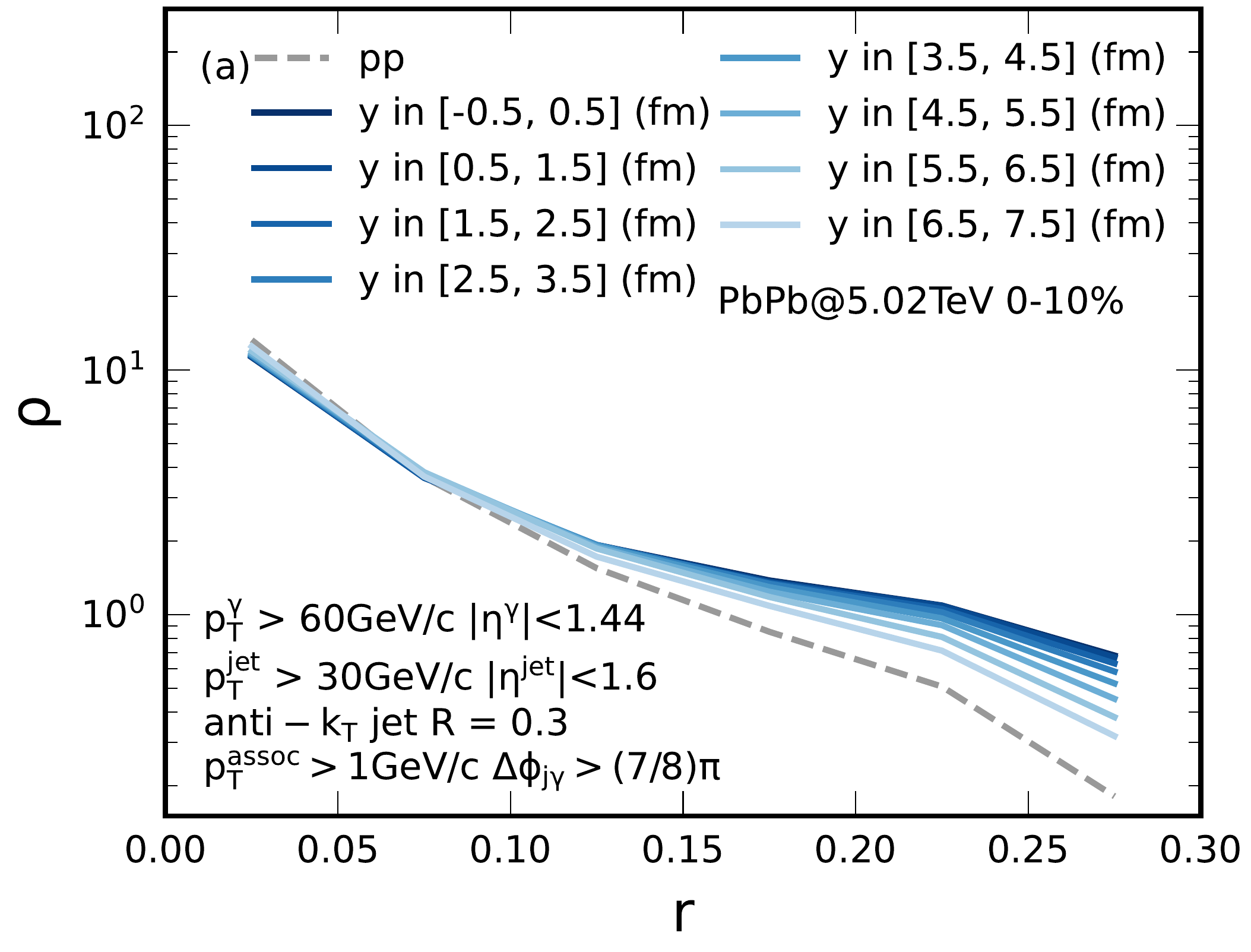}
\includegraphics[width=0.48\textwidth]{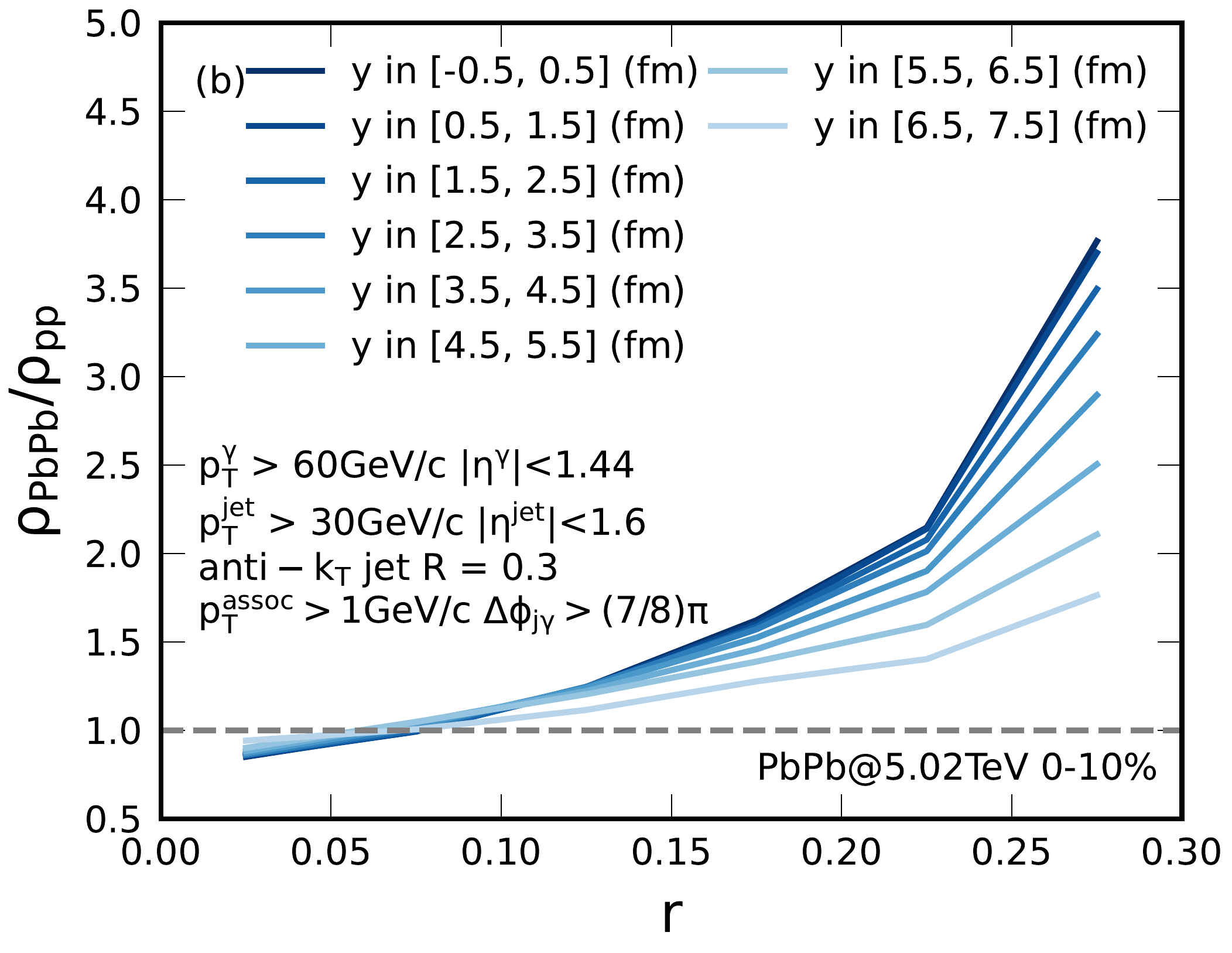}
\captionsetup{justification=raggedright}
\caption{(Color online) (a) Jet shape of $\gamma$-jets as a function of $r$ in 0-10\% Pb+Pb (solid curves) selected with different initial production position $y$ and p+p (dashing curve) collisions at $\sqrt{s_{NN}}=5.02$ TeV. (b) Ratios of jet shapes in Pb+Pb over p+p collisions as a function of $r$.}
\label{fix_position}
\end{figure}

For theoretical illustrations, we also show in Fig.~\ref{fix_position} the jet shape modification of $\gamma$-jets selected according to their transverse coordinates $y$ which shows similar trends for the corresponding $A_N^y$ dependence of the modification in Fig.~\ref{fix_asymmetry}. 

The broadening of jet shape is caused by interaction between the jet and the medium, in which hard partons at the core of a jet lose energy leading to a slight suppression of the jet shape at the core $r<0.05$. This lost energy will be carried by radiated gluons and medium-response to large angles, therefore transporting the energy towards the outer layer of the jet cone and leading to the enhancement of the jet shape in large $r$ regions. Selection of jet events with different values of the transverse asymmetry localizes the transverse positions of the initial jet production with different average path lengths and therefore different degree of jet quenching. This naturally leads to different levels of broadening of the jet shape, decreasing with increasing values of the transverse asymmetry or the average values of the transverse coordinate ($y$). 
This observation is also consistent with previous studies \cite{Luo:2018pto, Luo:2021hoo}.

\begin{figure}[htpb]
\raggedright
\includegraphics[width=0.48\textwidth]{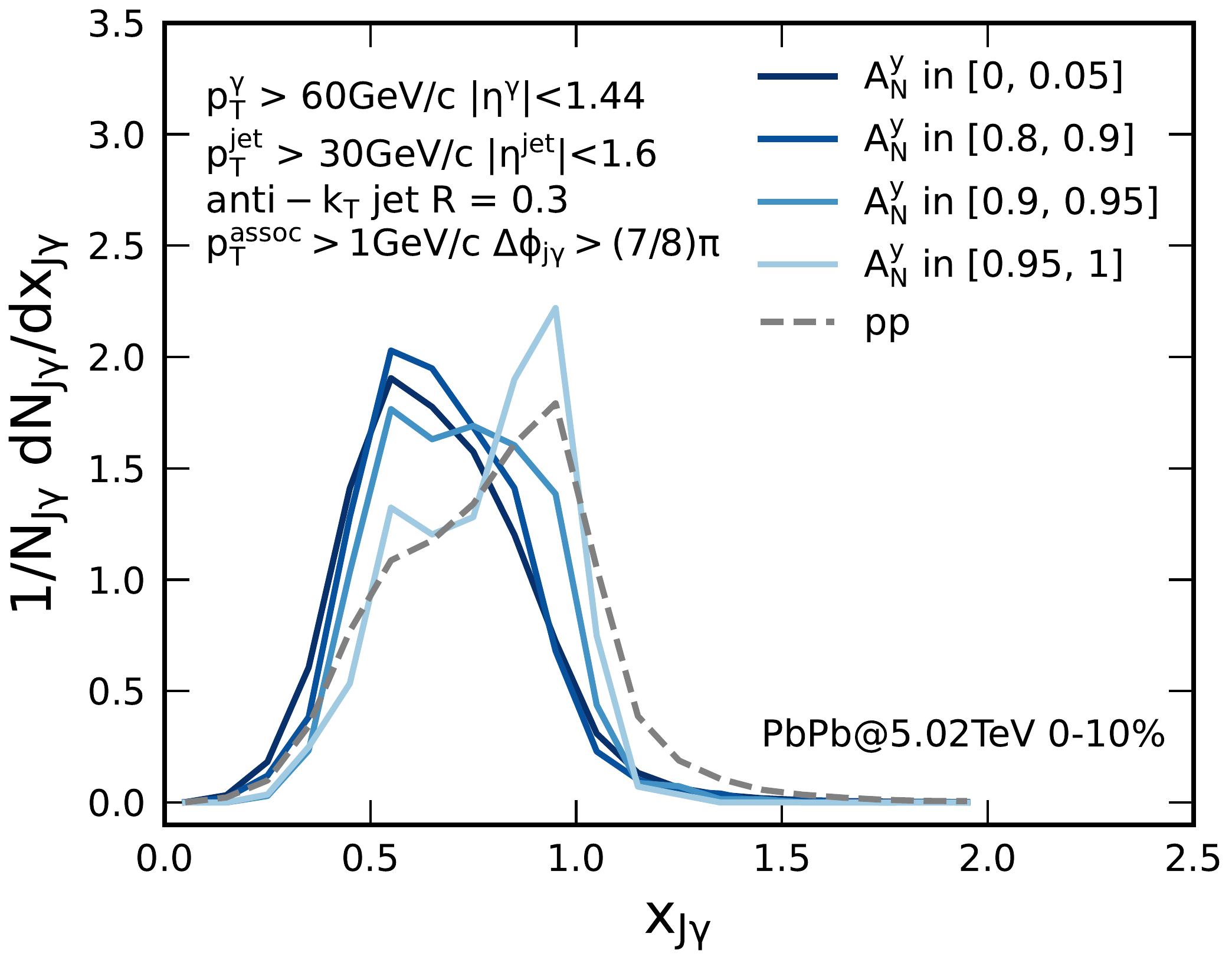}
\captionsetup{justification=raggedright}
\caption{(Color online) $x_{J\gamma}$ distribution, with $\gamma$-jets selected with different ranges of the transverse asymmetry $A_N^y$, in in 0-10\% Pb+Pb (solid line) and p+p (dashing line) collisions at $\sqrt{s_{NN}}=5.02$ TeV.}
\label{x_jg_A}
\end{figure}

\begin{figure}[htpb]
\raggedright
\includegraphics[width=0.48\textwidth]{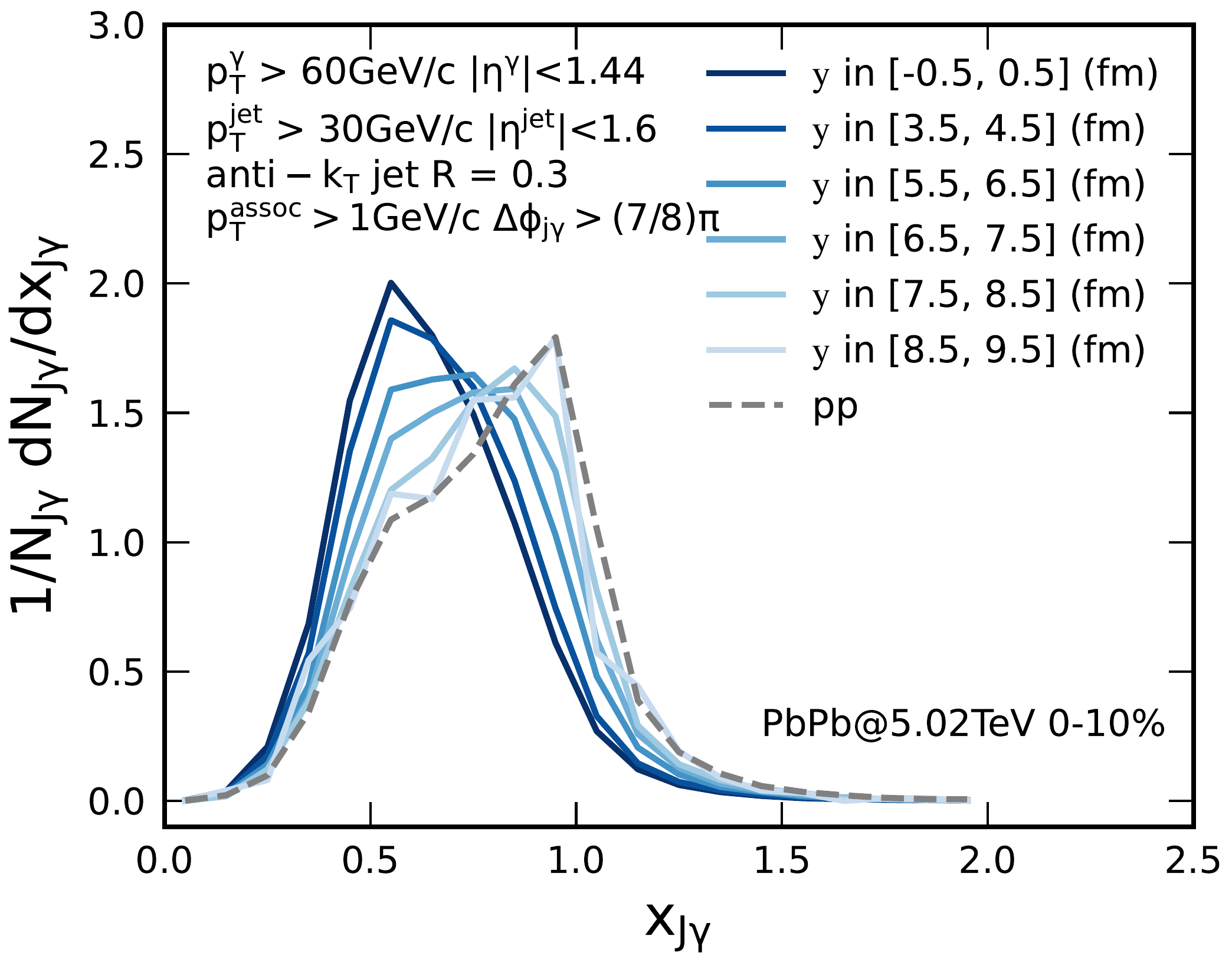}
\captionsetup{justification=raggedright}
\caption{(Color online) $x_{J\gamma}$ distribution, with $\gamma$-jets selected with different ranges of the transverse coordinates y, in 0-10\% Pb+Pb (solid line) and p+p (dashing line) collisions at $\sqrt{s_{NN}}=5.02$ TeV.}
\label{x_jg_y}
\end{figure}

Show in Fig. \ref{x_jg_A} is the $\gamma$-jet production rate as a function of $x_{J\gamma}$ ($=p_T^{\rm jet}/p_T^{\gamma}$), selected with different ranges of the transverse asymmetry $A_N^y$, in 0-10\% Pb+Pb at $\sqrt{s_{NN}}=5.02$ TeV. The dashed curve is the result from p+p collisions for comparison. One can see that the rate in Pb+Pb shifts toward smaller $x_{J\gamma}$ relative to p+p collisions due to jet quenching. The shift is smaller for $\gamma$-jets selected with a larger $A_N^y$ range, indicating more bias toward surface emission. For theoretical illustrations, we also show in Fig. \ref{x_jg_y} the $\gamma$-jet production rate as a function of $x_{J\gamma}$ with different initial jet production positions in the $y$ axis. The trend is similar to the corresponding dependence of $A_N^y$ as shown in Fig.~\ref{x_jg_A}.

\begin{figure}[htpb]
\raggedright
\includegraphics[width=0.49\textwidth]{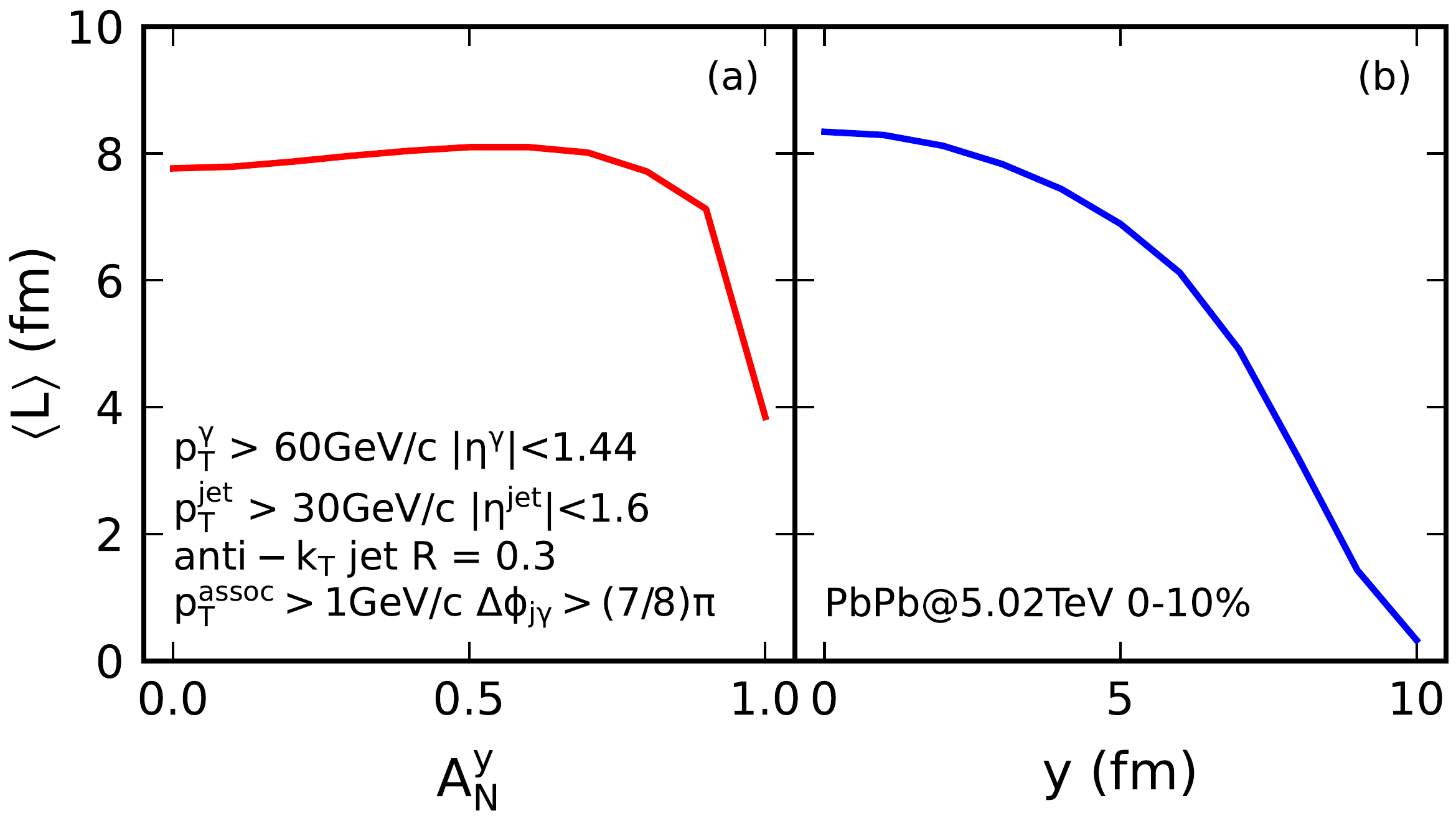}
\captionsetup{justification=raggedright}
\caption{(Color online) Average propagating path length of the leading jet partons as a function of $A_N^y$ (left) and as a function of initial production transverse position $y$ (right).}
\label{path_length}
\end{figure}

Although the $x_{J\gamma}$ distributions in Fig. \ref{x_jg_A} for different $A_N^y$ selections exhibit the similar trends as in Fig. \ref{x_jg_y} for the results with different selection of the initial production position in the $y$ axis, the $x_{J\gamma}$ distribution in Pb+Pb is observed to move back to the distribution in p+p collisions only when $A_N^y$ is sufficiently large. To understand these trends, we also show in Fig.~\ref{path_length} the average path length of the leading partons $\left\langle L \right\rangle$ which is defined as,
\begin{equation}
\left\langle L (A_N^y) \right\rangle = \frac{\int dxdy  P(x,y, A_N^y) l(x,y, A_N^y)}{\int dxdy P(x,y, A_N^y)},
\end{equation}
where $l(x,y,A_N^y)$ is the distance between the longitudinal position of the initial jet production and the boundary of QGP, and $P(x,y,A_N^y)$ is the initial jet production rates at $(x,y)$ for jet events with $A_N^y$. In the results shown in Fig. \ref{path_length}, we assume that the path is parallel to the $x$-axis and we take the critical temperature ($T_c=165$ MeV) at the boundary of the QGP.
The panel (a) in Fig. \ref{path_length} indicates that with increasing $A_N^y$,  $\left\langle L(A_N^y) \right\rangle$ remains nearly unchanged, but rapidly decreases as $A_N$ approaches to 1. Meanwhile, panel (b) shows that $\left\langle L(y) \right\rangle$ decreases smoothly with the initial jet production position $y$. Since the modification of the $\gamma$-jet  $x_{J\gamma}$ distributions is controlled mainly by jet energy loss which in turn is controlled by the jet path length, Fig. \ref{path_length} explains why modification of $x_{J\gamma}$ distribution is similar across most of the $A_N^y$ range where the average path length is insensitive to $A_N^y$. It reverts back to the distribution in p+p collisions only when $A_N^y$ is sufficiently large. 

\subsection{Jet shapes selected with longitudinal tomography}


In the previous subsection, we have seen how selecting different values of the transverse asymmetry $A_N^{\vec{n}}$ localizes the  initial transverse position ($y$) and leads to different jet propagation path length and jet shape broadening. Alternatively,  one can directly vary the jet propagation path length using longitudinal jet tomography, as illustrated in Fig.~\ref{pt_tomography}. 

\begin{figure}[htpb]
\raggedright
\includegraphics[width=0.48\textwidth]{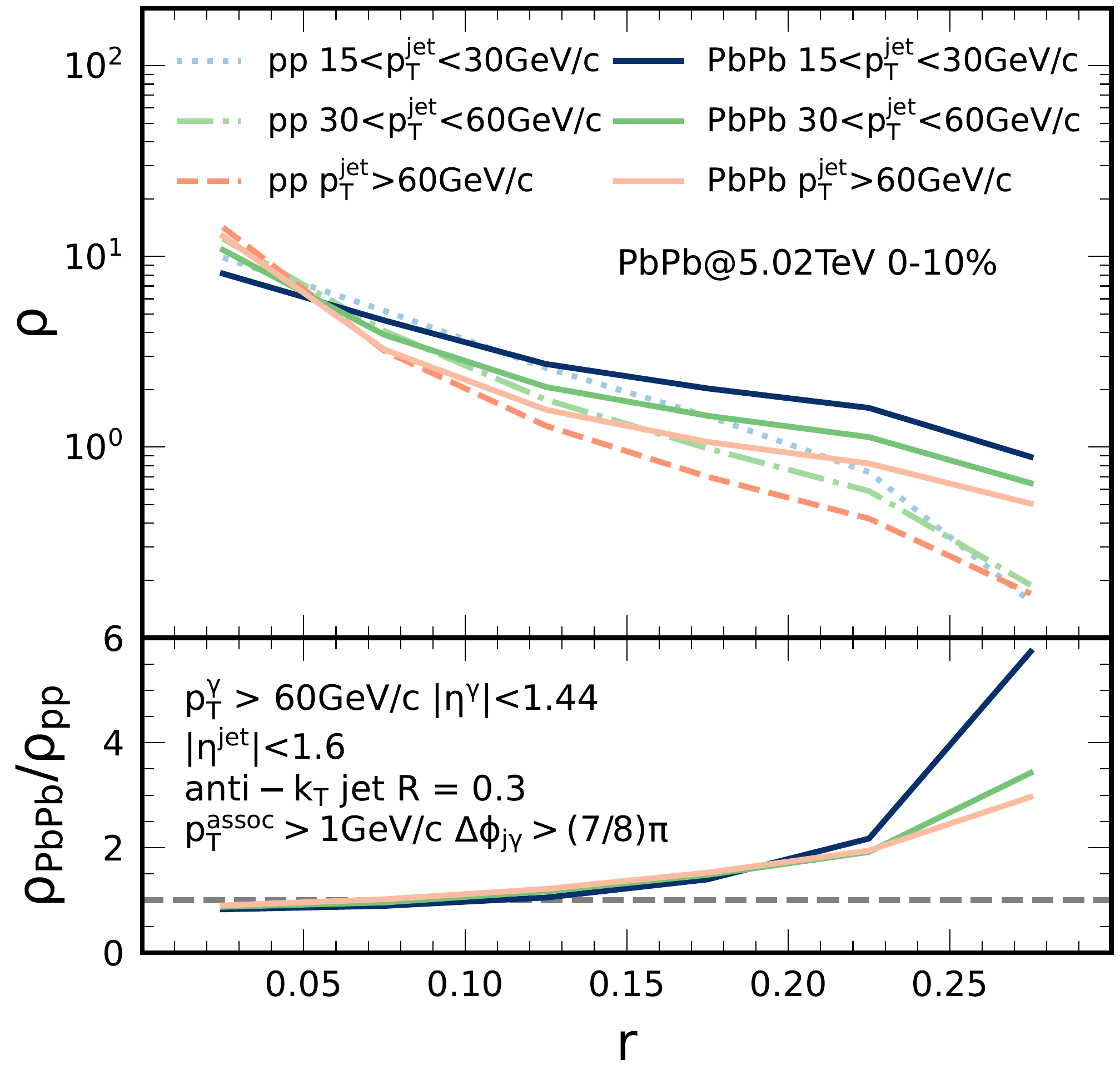}
\captionsetup{justification=raggedright}
\caption{(Color online) Upper panel: the jet shape as a function of $r$, selected with different jet transverse momentum, $p_T^{\rm jet} < 30$ GeV/c, $30< p_T^{\rm jet} < 60$ GeV/c and $p_T^{\rm jet} > 60$ GeV/c in in 0-10\% Pb+Pb and p+p collisions at $\sqrt{s_{NN}}=5.02$ TeV, respectively. The transverse momentum of the trigger photon is $p_T^\gamma>60$ GeV/c.
Lower panel: the ratio of jet shapes between Pb+Pb and p+p collisions. Other kinematic cuts are the same as in Fig.~\ref{fix_asymmetry}.}
\label{pt_trigger_shape}
\end{figure}

Shown in Fig.~\ref{pt_trigger_shape} are (a) the jet shapes as a function of $r$ for $\gamma$-jets in 0-10\% Pb+Pb and p+p collisions at $\sqrt{s_{NN}}=5.02$ TeV, with different $p_T^{\rm jet}$ and (b) the ratio of jet shapes between Pb+Pb and p+p collisions.  Because of QCD branching in vacuum, the jet shape in p+p becomes broader with larger jet transverse momentum $p_T^{\rm jet}$ as we see in the figure. This energy-dependence of the vacuum jet shape can lead to jet shape broadening due to jet energy loss in Pb+Pb collisions. In addition, jet-medium interaction will also broaden the jet shape because of radiated gluons and medium-response at large angles. This medium-induced jet shape broadening will also depend on $p_T^{\rm jet}$ because of the path-length bias imposed by the longitudinal jet tomography.

As we have seen in Fig.~\ref{pt_tomography}, for given values of the photon transverse momentum $p_T^\gamma$, $\gamma$-jets with small final $p_T^{\rm jet}$ normally comes from jets initially produced at the center of QGP that have gone through a along path length and lost significant amount of energy. One should also see large broadening of the jet shapes.   On the other hand,   $\gamma$-jets with large $p_T^{\rm jet}$ are more likely produced initially  close to the surface of QGP, travel a shorter path length and lose less energy. The corresponding jet shape broadening is also smaller as we see in Fig.~\ref{pt_trigger_shape}.

It is worth noting again that $\gamma$-jets with the same final $p_T^{\rm jet}$ in Pb+Pb collisions usually come from initial jets with higher transverse momentum than in p+p collisions due to jet energy loss. This energy loss alone can lead to jet shape broadening due to energy-dependence of the jet shape in vacuum. Such jet shape broadening due to jet energy-loss is approximately independent of the final jet energy $p_T^{\rm jet}$  as we see in the LBT results for $r<0.2$. At large radius $r>0.22$, the jet shape broadening due to radiated gluons and medium-response strongly depends on  $p_T^{\rm jet}$ due to path-length bias imposed by the longitudinal jet tomography.

\subsection{Jet shapes selected with 2D tomography}

One can combine transverse and longitudinal jet tomography, or 2D jet tomography to study the jet shape modification in Pb+Pb collisions  by simultaneously selecting both final jet $p_T^{\rm jet}$ and transverse asymmetry $A_N^y$.
We select trigger photons with $p_T^{\gamma} > 60$ GeV/$c$ along the negative $x$-axis direction. We calculate the jet shape of the correlated jets with two intervals of transverse momentum: small $p_T^{\rm jet} = [15, 30]$ GeV and large $p_T^{\rm jet} > 60$ GeV. In addition, we require the $\gamma$-jets have small $A_N^y$ = [-0.05, 0.05] and large transverse asymmetry $A_N^y$ = [0.55, 0.65]. 

\begin{figure}[htpb]
\raggedright
\includegraphics[width=0.48\textwidth]{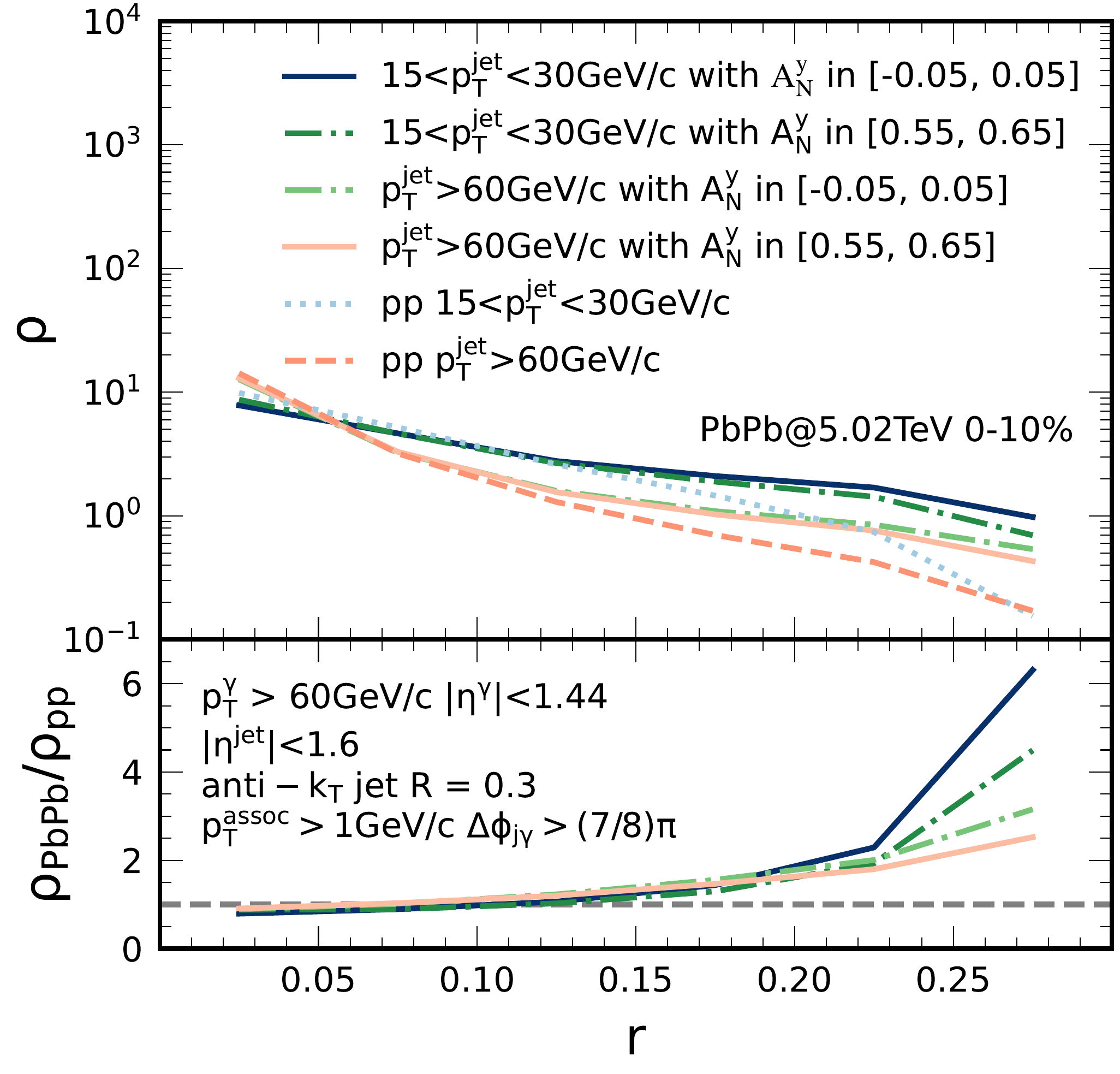}
\captionsetup{justification=raggedright}
\caption{(Color online) Similar to Fig. \ref{pt_trigger_shape}, but simultaneously selected with the two final observables, $p_T^{\rm jet}$ = [15, 30] GeV/c and $A_N^y$ = [-0.05, 0.05], $p_T^{\rm jet}$ = [15, 30] GeV/c and $A_N^y$ = [0.55, 0.65], $p_T^{\rm jet} > 60$ GeV/c and $A_N^y$ = [-0.05, 0.05], $p_T^{\rm jet} > 60$ GeV/c and $A_N^y$ = [0.55, 0.65], respectively.}
\label{largest_or_smallest}
\end{figure}

The jet shapes of these $\gamma$-jets and their corresponding ratios between 0-10\% Pb+Pb and p+p collisions at $\sqrt{s_{NN}}=5.02$ TeV are shown Fig. \ref{largest_or_smallest} for four different combinations of transverse and longitudinal tomographic selections. Among these four sets of tomographic selected configurations, $\gamma$-jets with small $p_T^{\rm jet}$ in conjunction with small $A_N^y$ have the most broadened jet shapes, since these jets are produced at the central region of QGP and traverse the longest path length on the average.
Conversely, the configurations that have the least jet shape broadening are those $\gamma$-jets with large values of both $p_T^{\rm jet}$ and $A_N^y$, that are initially produced in the outer region of QGP and travel a smaller path-length.  If the $\gamma$-jets are selected with small $p_T^{\rm jet}$ and large $A_N^y$, or large $p_T^{\rm jet}$ and small $A_N^y$, the medium modification of the jet shape falls between the above two extreme cases as seen in Figure \ref{largest_or_smallest}.

\section{Asymmetric jet shapes}

\begin{figure*}[htb]
\centering
\includegraphics[width=0.82\textwidth]{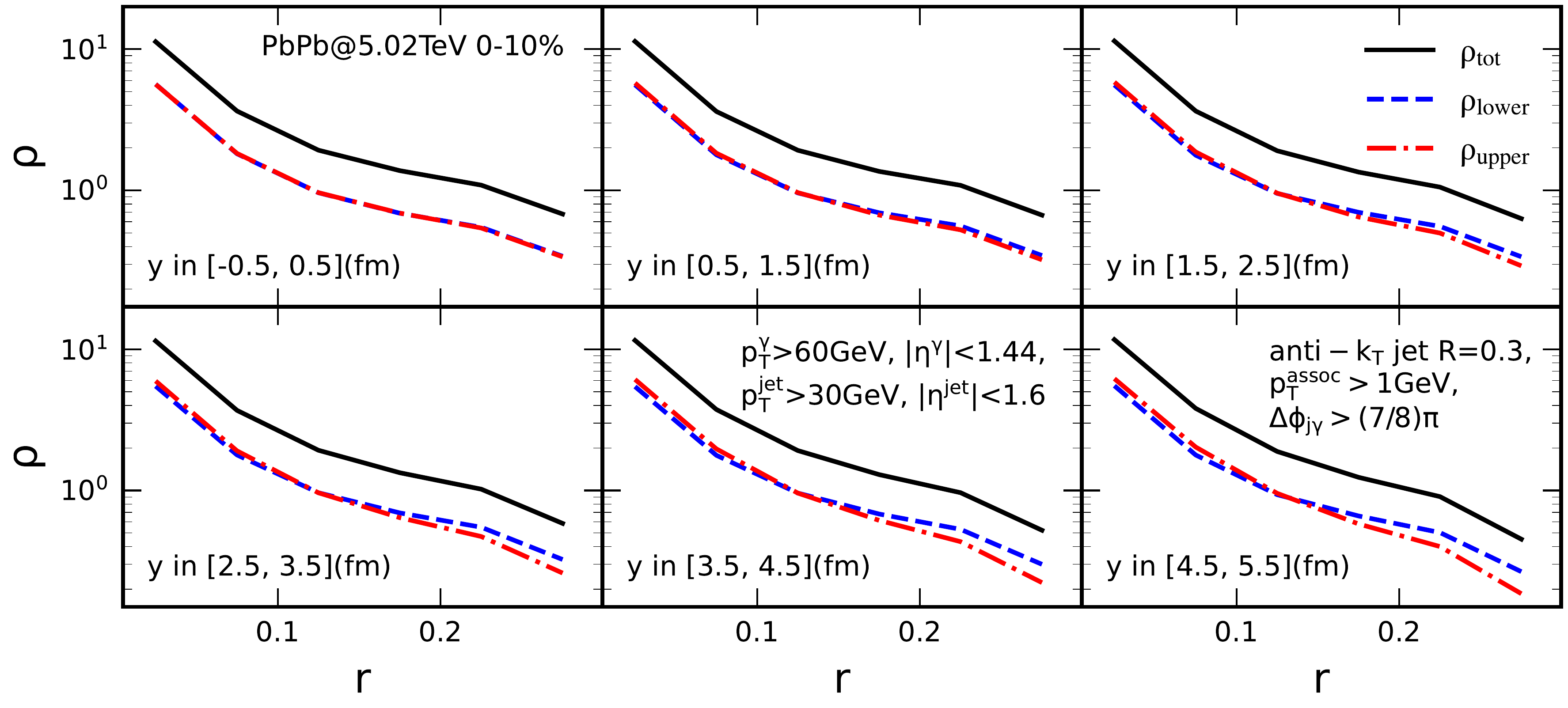}
\captionsetup{justification=raggedright}
\caption{(Color online) Comparisons of $\rho_{upper}$ (red dash-dotted) and $\rho_{lower}$ (blue dashed ) as a function of $r$ in 0-10\% Pb+Pb collisions at $\sqrt{s_{NN}}=5.02$ TeV, with different given $y$ positions of the initial jet production. Other cuts are the same as in Fig.~\ref{asymmetry}.}
\label{jet_asymmetry}
\end{figure*}

 The dependence of the jet shape modification on variables defined in different types of tomography methods in the last section is essentially the manifestation of the path-length dependence of jet-medium interaction and the associated medium response. We will study in this section the jet shape modification specifically caused by the gradient of the parton density or the jet transport coefficient $\hat q$ and the radial flow.  We will look at the azimuthal angle distribution of energy density inside the jet cone relative to the direction of the transverse gradient of the QGP medium. We consider the same $\gamma$-jet configurations relative to the event plane of heavy-ion collisions as illustrated in Fig~\ref{illustration}. Because of the gradient of the medium density and the radial flow, both final jet partons and medium-response distribution will be asymmetric relative to the direction of the gradient.

 For a purely theoretical illustration,  we first divide the jet cone into two equal halves using a plane defined by the beam direction and the direction of the jet axis as shown in Fig~\ref{illustration}. We then calculate the jet shapes as a function of the radius $r$ for each half cone - one for the upper half and another for the lower half,
\begin{align}
\rho(r)_{\rm upper/lower} = \frac{1}{\Delta r} \frac{\sum_i^{N_{\rm jet}} p_T^i (r-\frac{\Delta r}{2}, r+\frac{\Delta r}{2}) \Theta (\pm \vec{k}\cdot \vec{n})}{\sum_i^{N_{\rm jet}} p_T^i(0, R)},
\label{eqn:shape_half}
\end{align}
where $\pm$ in the $\Theta$ function selects associated particles in the upper half and lower half of the jet cone, respectively, $\vec{k}$ is the momentum of the associated particle and $\vec{n}$ is the norm of the plane defined by the beam direction and the momentum of this jet,  which coincides with the event plane in our setup.

Shown in Fig.~\ref{jet_asymmetry} are the upper (red dash-dotted), lower half (blue dashed), and the total jet shape (black solid line) for $\gamma$-jets initially produced in 6 different regions of the transverse coordinate $y$ in 0-10\% Pb+Pb collisions at $\sqrt{s_{NN}}=5.02$ TeV.
For jets produced in the central region of the QGP with $y$ = [-0.5, 0.5] fm, the upper and lower jet shapes are almost identical, while jets produced near the system edge with $y$ = [4.5, 5.5] fm have a broader lower-half jet shape as compared to the upper-half due to the asymmetric parton transport of jet partons and medium response in the non-uniform QGP medium.

\begin{figure}[htpb]
\raggedright
\includegraphics[width=0.48\textwidth]{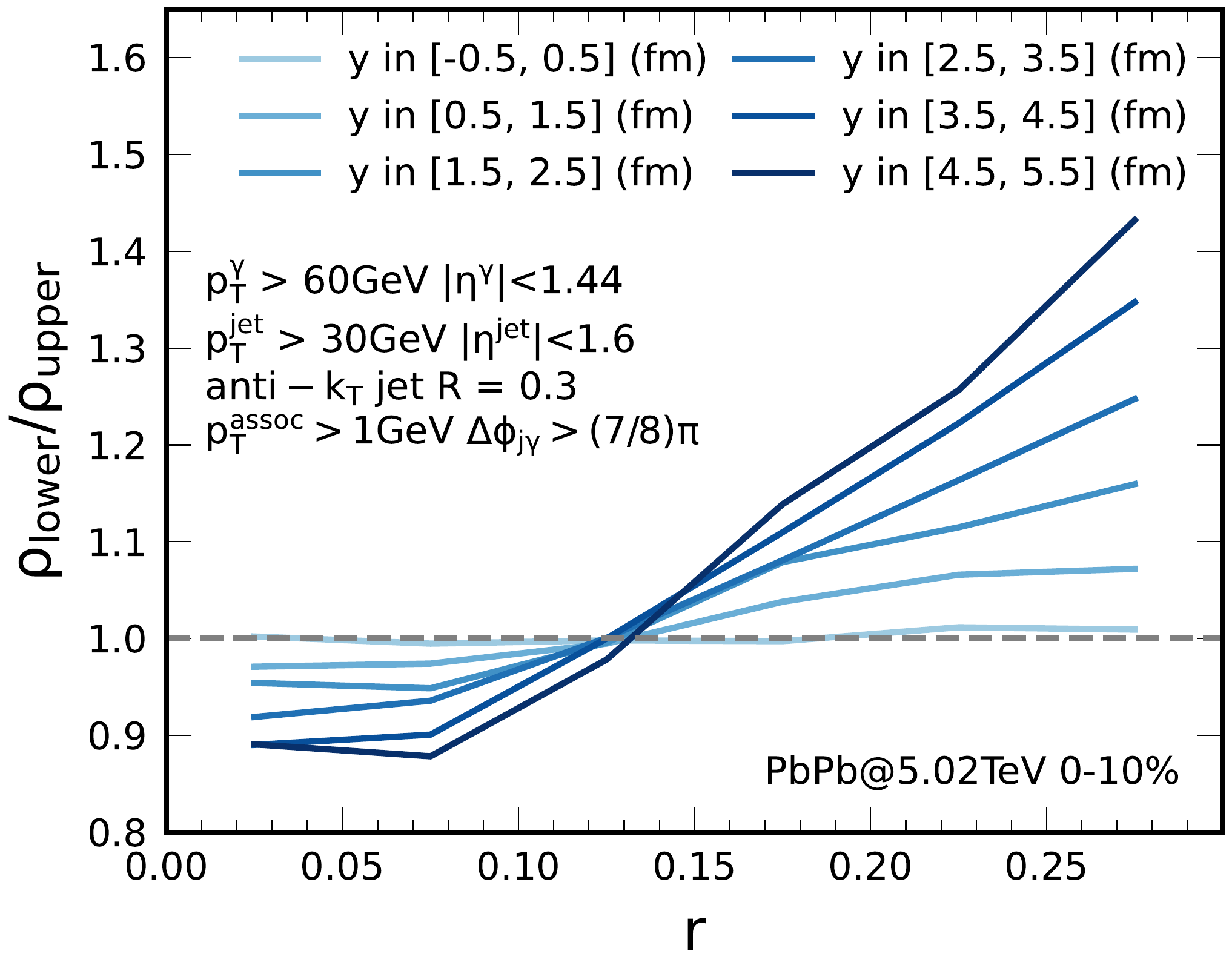}
\captionsetup{justification=raggedright}
\caption{(Color online) The ratio of $\rho_{\rm lower}$ over $\rho_{\rm upper}$ as a function of $r$. 
Kinematic selections and cuts are the same as in Fig.~\ref{jet_asymmetry}.}
\label{above_vs_under}
\end{figure}

 We also plot the ratio between $\rho_{\rm lower}$ and $\rho_{\rm upper}$ in Fig.~\ref{above_vs_under} as a function of the radius $r$ for jets initially produced in different regions of the QGP. It is clear that the lower-half jet shape is suppressed at the core of the jet cone but enhanced toward the edge of the jet cone as compared to the upper-half jet shape.  Since the jet core is made up of the energetic jet partons, their diffusion in the transverse direction is enhanced toward the dilute region of the QGP medium (opposite direction of the transverse gradient $d\hat q/dy$) leading to the positive value of the transverse asymmetry $A_N(y)$ and the suppression of $\rho_{\rm lower}$ relative to $\rho_{\rm upper}$ at the core of the jet cone. Jet-induced medium response in the same jet event will compensate such transport of momentum in the transverse direction and flow in the opposite direction of the transverse diffusion of hard jet partons (along the direction of the transverse gradient $d\hat q/dy$). This therefore contributes to the enhancement of the lower-half jet shape $\rho_{\rm lower}$ relative the upper-half $\rho_{\rm upper}$ in the large radius region where medium response dominate. The difference between $\rho_{\rm lower}$ and $\rho_{\rm upper}$ is more pronounced for jets produced in the outer region of the QGP where the gradient of $\hat{q}$ increases toward edge due to the initial energy density distribution of the hot matter.


\begin{figure}[htb]
\raggedright
\includegraphics[width=0.48\textwidth]{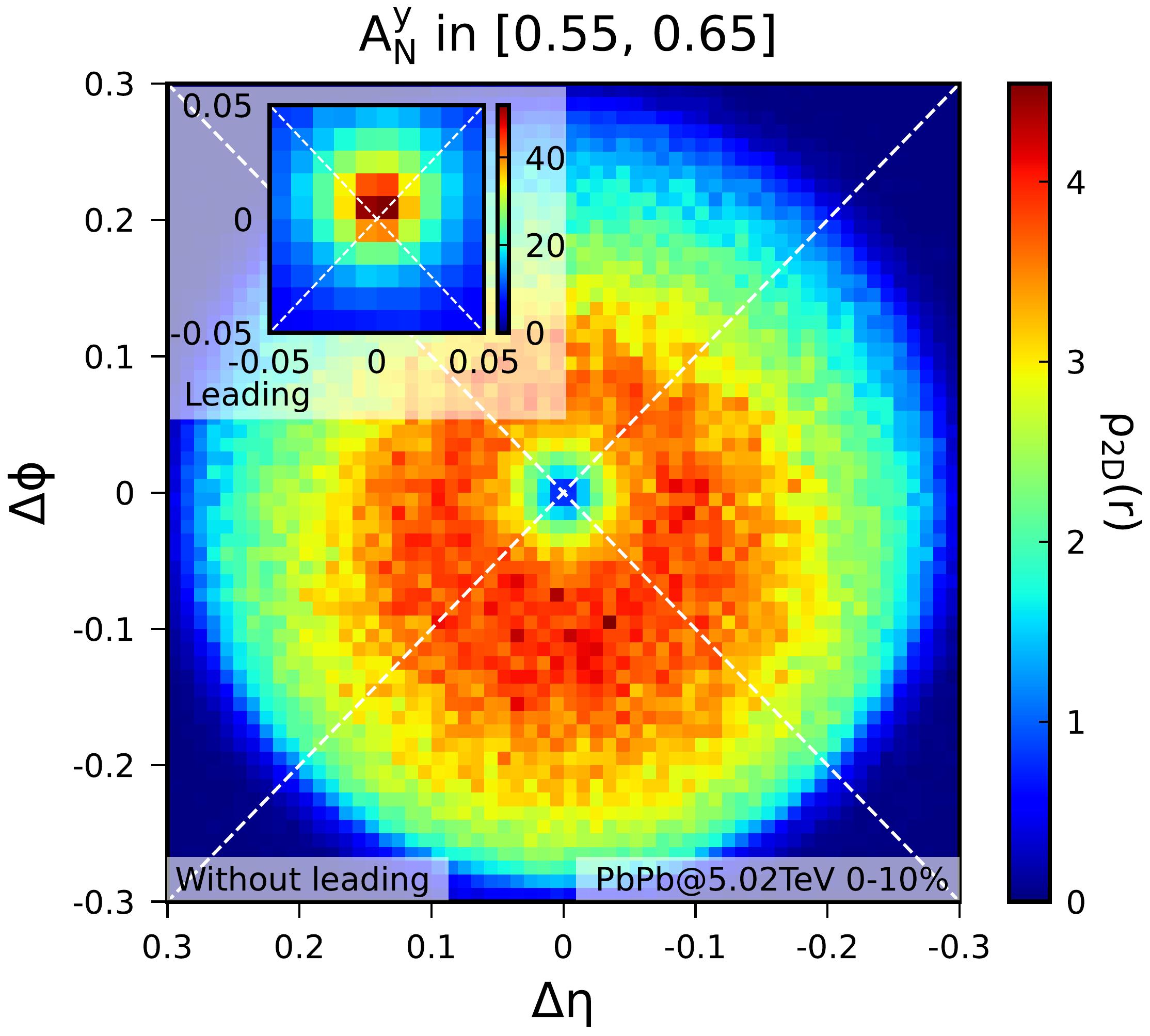}
\captionsetup{justification=raggedright}
\caption{(Color online) The in-cone transverse momentum distribution as a function of $\Delta\phi$ and $\Delta\eta$ in 0-10\% $Pb+Pb$ collisions at $\sqrt{s_{NN}}=5.02$ TeV with $A_N^y = [0.55, 0.65]$. Kinematic selections and cuts are the same as in Fig. \ref{asymmetry}.}
\label{phi_eta_soft_hard}
\end{figure}

Note that in experiments one can only statistically localize the transverse region of the initial jet production according to the value of the jet transverse asymmetry $A_N^y$. However, fluctuations of $A_N^y$ make it difficult to determine the transverse location precisely in each event, especially for small values of $A_N^y$ as seen in Fig.~\ref{A_distribution}.
In order to observe the asymmetric jet shape as we have illustrated in the about theoretical study, we introduce a two-dimensional $r$-weighted jet shape $\Tilde{\rho}_{2D}(\Delta \eta, \Delta \phi)$ for the transverse momentum distributions inside the jet cone in the $\Delta\eta-\Delta\phi$ plane,
 \begin{align}
     \Tilde{\rho}_{2D}(\Delta \eta, \Delta \phi) &= \frac{1}{\delta\eta \delta \phi} \nonumber\\
     &\frac{\sum_{\rm jet} \sum_{(\Delta\eta, \Delta\phi)}^{(\Delta\eta+\delta\eta, \Delta\phi+\delta\phi)}r_{i}p_T^{\rm i}/p_T^{\rm jet}}{\sum_{\rm jet}\sum_{r_{i}<R} r_{i} p_T^{\rm i}/p_T^{\rm jet}},
 \end{align}
 where $r_{i}=\sqrt{\Delta\eta^2+\Delta\phi^2}$ is the distance between the $i$-th associated parton and the jet axis, $p_T^{\rm i}$ is the transverse momentum of the $i$-th associated parton, $\Delta\eta=\eta_{i}-\eta_{jet}$ and $\Delta\phi=\phi_{i}-\phi_{jet}$, the $\delta\eta$ and $\delta\phi$ are the bin sizes in the $\Delta \eta$ and $\Delta \phi$ directions, respectively. The 2-D jet profile is weighted with the distance $r$ to enhance the contribution of soft partons in the large $r$ region from medium response.

The 2D jet shape $\Tilde{\rho}_{2D}(\Delta \eta, \Delta \phi)$ inside the jet cone is shown in Fig.~\ref{phi_eta_soft_hard}, for $\gamma$-jets in 0-10\% Pb+Pb collisions at $5.02$ TeV, with $A_N^y$ in the range $[0.55, 0.65]$. The contribution from the leading hard jet parton to the 2-D jet profile at the core is shown in the inset, while the remaining contribution (with the leading parton subtracted) primarily from soft particles to the 2D jet profile inside the jet cone is displayed in the main figure. 
This 2D jet shape with given transverse asymmetry indeed shows the same asymmetrical structure as we have discussed in the above study with specified region of transverse position of the initial jet production. The hard jet partons are diffused toward the dilute region of the QGP  medium while soft partons from the medium response tend to move to the direction of the dense region of the QGP. 
In other words, the asymmetric jet shape suggests that hard partons inside the jet are deflected away from dense regions, while soft partons from medium response are more likely to be created in dense regions. This asymmetrical behavior is particularly more pronounced for $\gamma$-jet events with large $A_N^y$, where the jet is produced mostly in the outer layer of the medium, and is tangential to the surface.

\begin{figure}[htb]
\raggedright
\includegraphics[width=0.46\textwidth]{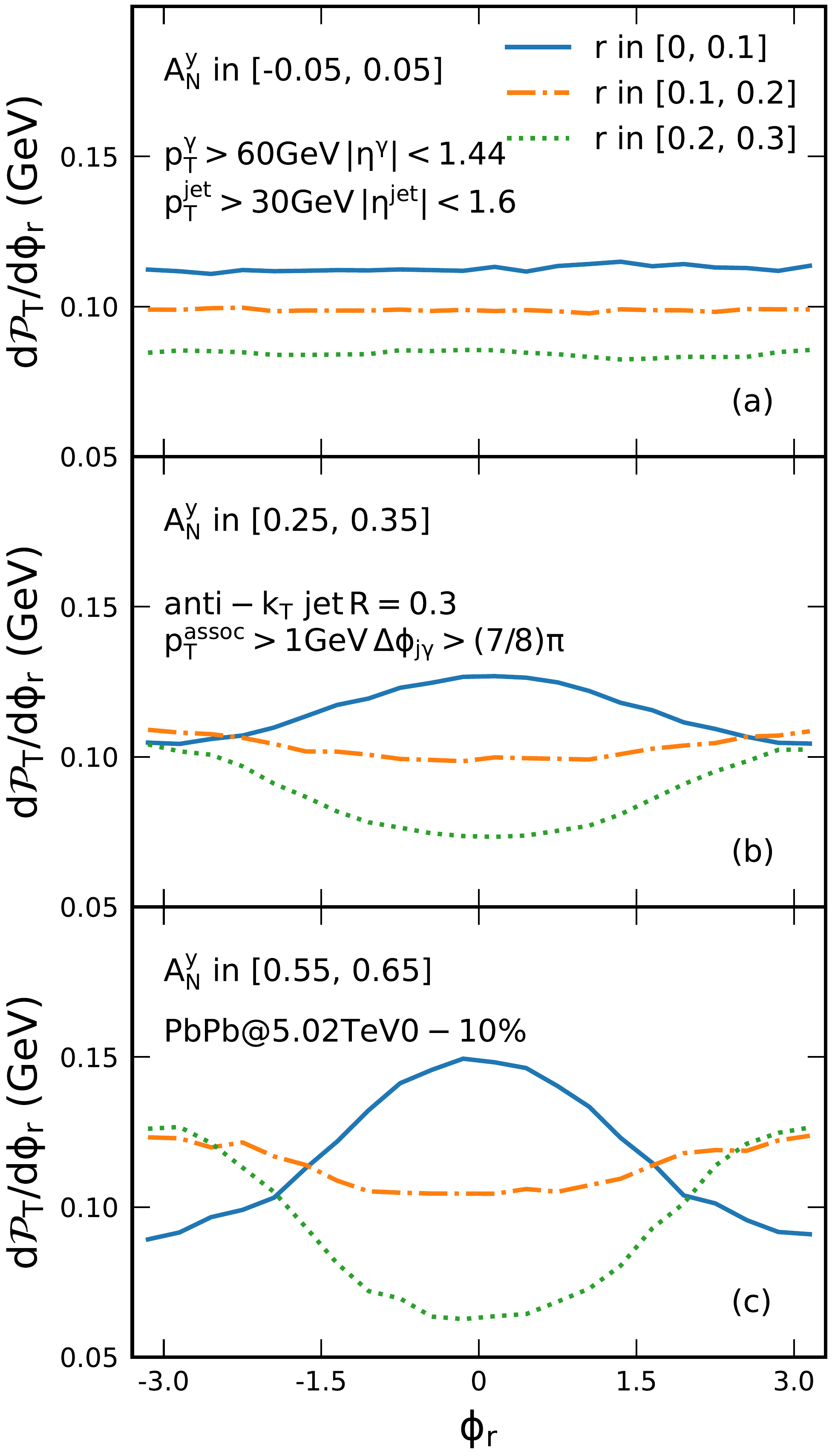}
\captionsetup{justification=raggedright}
\caption{(Color online) The in-cone transverse momentum distribution as a function of $\phi_r$ in three different circular regions inside $\gamma$-jets in 0-10\% $Pb+Pb$ collisions at $\sqrt{s_{NN}}=5.02$ TeV. Three kinds of jet events are selected with different $A_N^y$ values in panels (a), (b), and (c), respectively. 
}
\label{phi_y_dist}
\end{figure}

\begin{figure}[htb]
\raggedright
\includegraphics[width=0.46\textwidth]{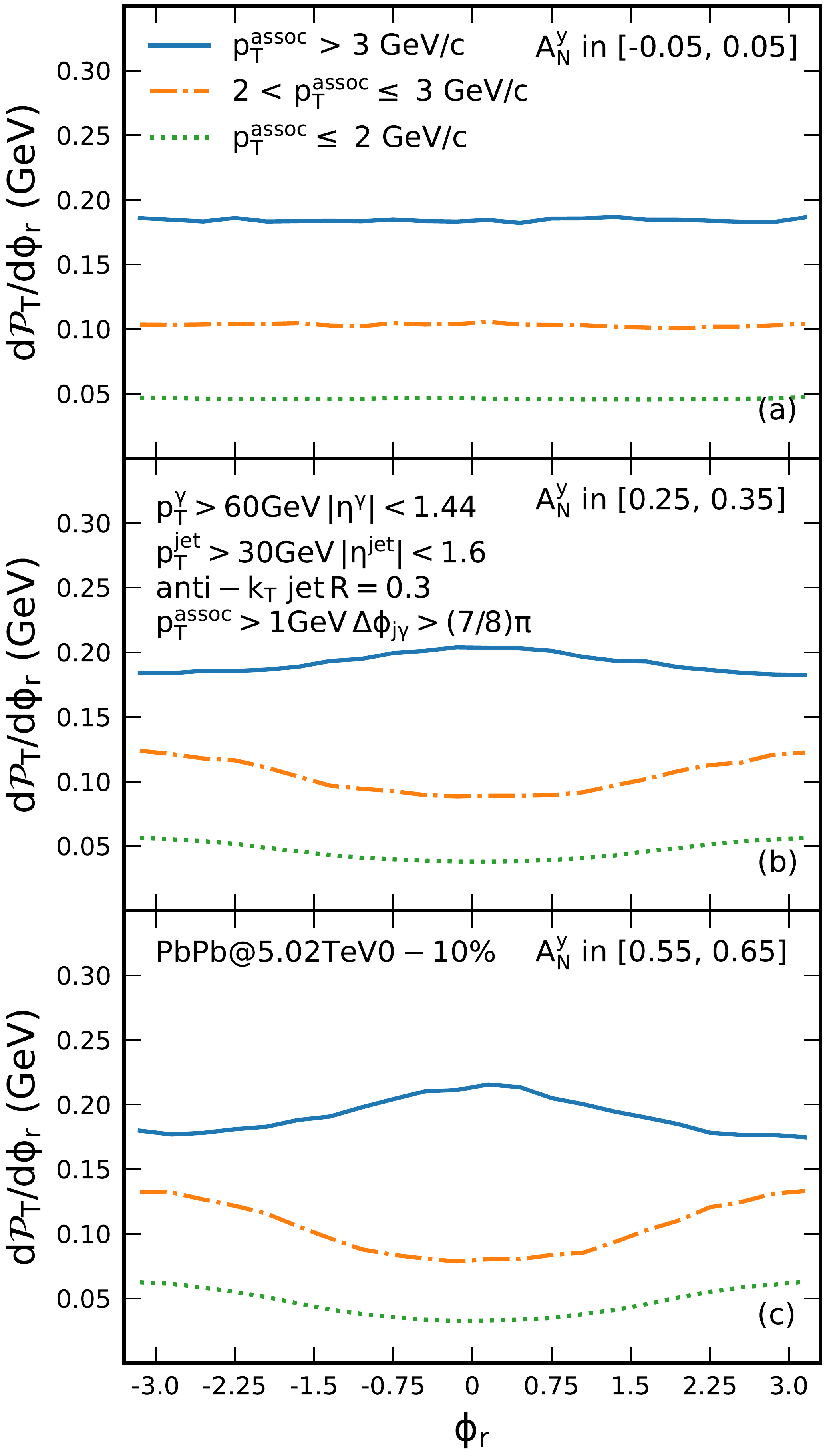}
\captionsetup{justification=raggedright}
\caption{(Color online) The in-cone transverse momentum distribution as a function of $\phi_r$ in three different $p_T^{\rm assoc}$ ranges in 0-10\% $Pb+Pb$ collisions at $\sqrt{s_{NN}}=5.02$ TeV. Three kinds of jet events are selected with different $A_N^y$ values in panels (a), (b), and (c), respectively. 
}
\label{phi_y_dist_2}
\end{figure}


 To further illustrate the asymmetrical jet shape and the underlying contributions from hard and soft particles, we project the 2D jet shape to the azimuthal angle distribution inside the jet cone,
\begin{align}
\frac{d{\cal P}_T}{d\phi_r} = \frac{1}{N_{\rm jet}} \frac{\sum_{\rm jet} \sum_{\phi_r}^{\phi_r+\Delta\phi_r}r_{i} \cdot p_T^{\rm i}}{\Delta\phi_r},
 \end{align}
where $N_{\rm jet}$ is the total number of jet events, and $\phi_r$ is the azimuthal angle in the $\Delta \eta - \Delta \phi$ plane,
\begin{align}
\begin{cases}
\begin{aligned}
\phi_r &= \arcsin{\left(\frac{\Delta\eta}{r}\right)}, & (\Delta\phi \geqslant 0) \\[0.5em]
\phi_r &= \pi - \arcsin{\left(\frac{\Delta\eta}{r}\right)}, & (\Delta\phi < 0, \Delta\eta \geqslant 0) \\[0.5em]
\phi_r &= -\pi - \arcsin{\left(\frac{\Delta\eta}{r}\right)}, & (\Delta\phi < 0, \Delta\eta <0).
\end{aligned}
\end{cases}
\end{align}
with $\phi_r = 0$ defined as the direction of $y$ or along the positive direction of $A_N^y$. 

Shown in Fig. \ref{phi_y_dist} are the azimuthal angle $\phi_r$ distribution of transverse energy inside $\gamma$-jets in 0-10\% Pb+Pb collisions at $\sqrt{s_{NN}}=5.02$ TeV with three different jet transverse asymmetry (a)  $A_N^y = [-0.05, 0.05]$, (b) $[0.25, 0.35]$ (c) and $[0.55, 0.65]$ for three different circular annulus at radius $r$ = [0, 0.1] (solid), [0.1, 0.2] (dashed) and [0.2, 0.3] (dotted), respectively. 
We also show in Fig.~\ref{phi_y_dist_2} the angular $\phi_r$ distribution of transverse energy inside $\gamma$-jets with three different jet transverse asymmetry (a)  $A_N^y = [-0.05, 0.05]$, (b) $[0.25, 0.35]$ (c) and $[0.55, 0.65]$, for three different regions of parton transverse momentum, $p_T^{\rm assoc} > 3$ GeV/c (solid), $2 < p_T^{\rm assoc} \leq 3$ GeV/c (dashed) and $p_T^{\rm assoc} \leq 2$ GeV/c (dotted), respectively.  Soft partons originating from the medium response have smaller $p_T^{\rm assoc}$, while large $p_T^{\rm assoc}$ is associated with hard partons. We observe different trends in the $\phi_r$ distributions for hard and soft partons, particularly  for jet events with a large value of $A_N^y$. This conclusion is consistent with the results presented in Fig.~\ref{phi_y_dist}, except that different circular annuli are used to separate hard from soft partons:  hard jet partons are concentrated around the core of the jet cone ($r=0$) whereas soft partons tend  to dominate the region toward the edge of the jet cone.

With small jet transverse asymmetry $A_N^y$ ranging from -0.05 to 0.05, the jet shape is mostly symmetric with an approximately uniform distribution in $\phi_r$ at both small and large radius $r$ or for both small and large transverse momentum partons as seen in Figs.~\ref{phi_y_dist}(a) and \ref{phi_y_dist_2}(a). 
For jets with sizable or large transverse asymmetry $A_N^y$ in ranges  [0.25, 0.35]  or [0.55, 0.65],  they propagate through regions of QGP with large transverse gradient. These jets have asymmetric jet shapes with non-uniform azimuthal $\phi_r$ distributions as seen in Figs.~\ref{phi_y_dist}(b-c) and \ref{phi_y_dist_2}(b-c).  Hard jet partons at the core of the jet have a peak in the opposite direction of the gradient ($\phi_r=0$ toward the dilute region) while soft partons at large radius have a peak in the direction of the gradient ($\phi_r=\pi$ toward the dense region).
This is consistent with the findings as illustrated in Fig.~\ref{phi_eta_soft_hard}.


Note that in experimental measurements, one can only know the absolute value of the transverse asymmetry $|A_N^y|$ with equal probability of positive and negative values. If one averages events with both positive and negative values of $A_N^y$, the final azimuthal distribution will be uniform. To circumvent this, one should align the orientation of the hard jet (or soft) partons in each event before averaging over events.  The anti-correlation of hard and soft partons in asymmetric jets serves to determine the direction of the gradient of the QGP through which the jets propagate. This is an especially useful technique to measure the asymmetric jet shape caused by the transverse gradient of the QGP medium in central heavy-ion collisions where one cannot determine the event plane from the bulk hadron distributions.

\section{summary}

 We have investigated the medium-modified jet shapes of $\gamma$-jets in high-energy heavy-ion collisions using a two-dimensional jet tomography that combines gradient (transverse) and longitudinal jet tomography using event-by-event linear Boltzmann Transport simulations which can reproduce CMS experimental data. Transverse and longitudinal jet tomography are demonstrated to be able to localize the initial $\gamma$-jet production locations  as proposed in \cite{Zhang:2009rn} and \cite{He:2020iow}. By employing the 2D jet tomography in which  $\gamma$-jets are selected with different jet transverse asymmetry $A_N^{\vec n}$ or the longitudinal asymmetry $p_T^{\rm jet}/p_T^\gamma$, we computed the medium modification for the transverse momentum distribution inside the jet cone, namely the jet shape.  Jet shapes are in general broadened in Pb+Pb collisions relative to p+p collisions because of the energy transfer carried by soft partons in medium response at large angles relative to the jet axis.
 Our results indicate that  jet shape broadening are enhanced  in events with small values of $A_N^{\vec n}$ and $p_T^{\rm jet}$, corresponding to jets initially produced at the center of the QGP or with  long path lengths.  We further illustrated the asymmetrical jet shapes of these jets that propagate through regions of the QGP with finite transverse gradient. Hard jet partons at the core of the jet are defected away from the dense region of QGP while soft partons from the medium response are produced more in the dense region of QGP. This anti-correlation of hard and soft partons within the asymmetric jets can be used to determine the direction of the transverse gradient of the QGP in each jet event. These distinctive behaviors of hard and soft partons  inside the jet cone may pave the way for new avenues to investigate the properties of the QGP and strong interactions. 
 Future experimental measurements can validate these distinctive features in the jet shape, thereby testing the theory of QCD and parton transport.
 

\section*{Acknowledgments}

The authors would like to thank Guang-You Qin, Enke Wang and Zhong Yang for stimulating discussions. This work is supported by National Natural Science Foundation of China under Grants Nos. 11935007, 11435004, 12075098, 12305140, 12147134, by Guangdong Basic and Applied Basic Research Foundation No. 2021A1515110817, by Guangdong Major Project of Basic and Applied Basic Research No. 2020B0301030008, Science and Technology Program of Guangzhou No. 2019050001, and the Director, Office of Energy Research, Office of High Energy and Nuclear Physics, Division of Nuclear Physics, of the U.S. Department of Energy under Grant No. DE-AC02-05CH11231 and within the framework of the SURGE Collaboration, the US National Science Foundation under Grant No. OAC-2004571 within the X-SCAPE Collaboration.

\bibliography{citation}

\end{document}